\documentclass[aps,prc,twocolumn,floatfix,showpacs,amsmath,amssymb]{revtex4}
\usepackage{graphicx}
\usepackage{dcolumn}
\usepackage{color} 
  \def\nuc#1#2{\relax\ifmmode{}^{#1}{\protect\text{#2}}\else${}^{#1}$#2\fi}
  \def\itnuc#1#2{\setbox\@tempboxa=\hbox{\scriptsize\it #1}
    \def\@tempa{{}^{\box\@tempboxa}\!\protect\text{\it #2}}\relax
    \ifmmode \@tempa \else $\@tempa$\fi}
  \newcommand{\beel}{\nuc{11}{Be}}
  \newcommand{\beni}{\nuc{9}{Be}}
  \newcommand{\nm}{\ensuremath{N_\mathrm{max}}}
  \newcommand{\ho}{\ensuremath{\hbar \Omega}}
  \newcommand{\nn}{\ensuremath{N\!N}}
  \newcommand{\co}{(Color online)}
  \newcommand{\jpt}[4]{\ensuremath{\left( \frac{#1}{2}_{_#2}^{#3} \:
      \frac{#4}{2} \right)}}
  
 \newcommand{\avp}{AV8$^\prime$}
 \newcommand{\cdb}{CDB2k}
 \newcommand{\nlo}{N$^3$LO}
 \newcommand{\inoy}{INOY}
 
\begin{document}
\preprint{UCRL-JRNL-208555}
\title{Large basis \emph{ab initio} shell model investigation of
  $^{9}${Be} and $^{11}${Be}}
\author{C. Forss\'en}
\email[]{c.forssen@llnl.gov}
\author{P. Navr\'atil}
\author{W.E. Ormand}
\affiliation{Lawrence Livermore National Laboratory, P.O. Box 808, L-414, 
Livermore, CA  94551}
\author{E. Caurier}
\affiliation{Institut de Recherches Subatomiques
            (IN2P3-CNRS-Universit\'e Louis Pasteur)\\
            Batiment 27/1,
            67037 Strasbourg Cedex 2, France}

\date{\today}

\begin{abstract}
We are presenting the first \emph{ab initio} structure investigation of
the loosely bound $^{11}${Be} nucleus, together with a study of the
lighter isotope $^{9}${Be}. The nuclear structure of these isotopes is
particularly interesting due to the appearance of a parity-inverted
ground state in $^{11}${Be}. Our study is performed in the framework of
the \emph{ab initio} no-core shell model. Results obtained using four
different, high-precision two-nucleon interactions, in model spaces up
to 9$\hbar\Omega$, are shown. For both nuclei, and all potentials, we
reach convergence in the level ordering of positive- and negative-parity
spectra separately. Concerning their relative position, the
positive-parity states are always too high in excitation energy, but a
fast drop with respect to the negative-parity spectrum is observed when
the model space is increased. This behavior is most dramatic for
$^{11}${Be}. In the largest model space we were able to reach, the $1/2^+$
level has dropped down to become either the first or the second excited
state, depending on which interaction we use. We also observe a
contrasting behavior in the convergence patterns for different
two-nucleon potentials, and argue that a three-nucleon interaction is
needed to explain the parity inversion. Furthermore, large-basis
calculations of $^{13}${C} and $^{11}${B} are performed. This allows
us to study the systematics of the position of the first
unnatural-parity state in the $N=7$ isotone and the $A=11$ isobar. The
$^{11}${B} run in the $9\hbar\Omega$ model space involves a matrix with
dimension exceeding $1.1 \times 10^9$, and is our largest calculation so
far. We present results on binding energies, excitation spectra, level
configurations, radii, electromagnetic observables, and $^{10}\mathrm{Be} +
n$ overlap functions.
\end{abstract}
\maketitle
%
\section{\label{sec:intro}Introduction}
Studies of how nuclear structure evolves when varying the $N/Z$ ratio
are important in order to improve our fundamental understanding of
nuclear forces. For this reason, research on light neutron-rich nuclei
has attracted an increasing amount of theoretical and experimental
effort since the advent of radioactive nuclear beams. The application of
a standard mean-field picture to describe these few-body systems is
questionable, and it is not surprising that substantial deviations from
regular shell structure has been observed. The $A=11$ isobar is of
particular interest in this respect since it exhibits some anomalous
features that are not easily explained in a simple shell-model
framework. Most importantly, the parity-inverted $1/2^+$ ground state of
\beel\ was noticed by Talmi and Unna~\cite{tal60:4} already in the early
1960s, and it still remains one of the best examples of the
disappearance of the $N=8$ magic number.

Many theoretical studies of odd-$A$ beryllium isotopes have already
been performed using various models. A thourough review of the
structure of unstable light nuclei in terms of the shell model can be
found in Ref.~\cite{mil01:693}. Of particular interest is the study on
unnatural-parity states of the $A=11$ isobar by Teeters and
Kurath~\cite{tee77:275} using a 1\ho\ model space and the
Millener-Kurath interaction with modified $0s$ and $sd$
single-particle energies. The halo structure of the \beel\ ground
state was reproduced with the variational shell model by Otsuka
\emph{et al}~\cite{ots93:70}. They used Skyrme interactions and
constructed multi-nucleon wave functions from a variational
single-particle basis in a (0--1)\ho\ model space. Alternatively, the
loosely-bound nature of the valence neutron in \beel\ can be treated
explicitly in a $\nuc{10}{Be} + n$ picture with a Woods-Saxon
potential, see e.g.  Refs.~\cite{esb95:51,nun96:596}. Using a
coupled-channels treatment, the authors of these papers found a
significant overlap with excited-core states. Possible explanations
for the parity inversion of the \beel\ ground state has also been
investigated using the AMD+HF model~\cite{dot00:103}, which is a
combination of anti-symmetrized molecular dynamics with the concept of
single-particle motion. An extended version of the AMD framework was
later used to study excited states of \beel, and the existence of
three negative-parity rotational bands was proposed~\cite{kan02:66}.

There are also several calculations involving different cluster
models. In particular, $\alpha$-clustering has been considered to play
an important role in these systems. With this assumption as a starting
point, K. Arai \emph{et al} used an $\alpha + \alpha + n$ model and
obtained the ground state of \beni\ using the stochastical variational
method, while several particle-unbound excited states were studied
simultaneously with the complex scaling method~\cite{ara96:54}. A
similar $\alpha + \alpha + Xn$ description was employed by
P. Descouvemont~\cite{des02:699} in his study of possible rotational
bands in \nuc{9-11}{Be} using the Generator Coordinate Method. His
conclusion, however, was that the degree of $\alpha$-clustering
decreased with increasing mass, and consequently his model was not able
to reproduce some of the anomalous properties of \beel.

Unfortunately, due to the complexity of the problem, there has been no
genuine \emph{ab initio} investigation of \beel\ starting from realistic
inter-nucleon interactions. There is no doubt that the cluster and
potential models are very successful, and can provide reasonable
explanations for many observations. Still, one has to remember that they
rely upon the fundamental approximation that the total wave function can
be separated into cluster components. Furthermore, the effective
interactions used in all models must be fitted to some observables for
each individual case. On the contrary, a truly microscopic theory such
as the Green's Function Monte Carlo (GFMC) method~\cite{pud97:56}, or
the \emph{ab initio} no-core shell model
(NCSM)~\cite{nav00:84,nav00:62}, starts from a realistic inter-nucleon
potential and solves the $A$-body problem, producing an antisymmetrized
total wave function. It is a true challenge of our understanding of
atomic nuclei to investigate nuclear many-body systems ($A > 4$) using
such \emph{ab initio} approaches.

This paper represents an effort to fill this gap. Our study is performed
in the framework of the \emph{ab initio} NCSM, in which the $A$-body
Schr\"odinger equation is solved using a large Slater determinant
harmonic oscillator (HO) basis. However, it is well known that the HO
basis functions have incorrect asymptotics which might be a problem when
trying to describe loosely bound systems. Therefore it is desirable to
include as many terms as possible in the expansion of the total wave
function. By restricting our study to two-nucleon (\nn) interactions,
even though the NCSM allows for the inclusion of three-body forces, we
are able to maximize the model space and to better observe the
convergence of our results.

In Sec.~\ref{sec:theory} the framework for the NCSM will be briefly
outlined, and the four different high-precision \nn\ interactions that
are used in this work will be introduced. Sec.~\ref{sec:results} is
devoted to a presentation and discussion of our complete set of results
for \nuc{9,11}{Be}, with a particular focus on the position of the first
unnatural-parity state. Concluding remarks are presented in
Sec.~\ref{sec:conc}.
%
\section{\label{sec:theory}\emph{Ab initio} no-core shell model}
Applying the \emph{ab initio} NCSM is a multi-step process. The first
step is to derive the effective interaction from the underlying
inter-nucleon forces, and to transform it from relative coordinates into
the single-particle $M$-scheme basis. The second step is to evaluate and
diagonalize the effective Hamiltonian in an $A$-nucleon ($Z$ protons and
$N$ neutrons) Slater determinant HO basis that spans the complete $\nm
\ho$ model space. Finally, we can use the resulting wave functions for
further processing. This section contains a short discussion on each of
these steps. We stress that an important strength of the method is the
possibility to include virtually any type of inter-nucleon
potential. The four high-precision \nn\ interactions that have been used
in this study will be introduced in Sec.~\ref{sec:int}. A more detailed
description of the NCSM approach, as it is implemented in this study,
can be found in, e.g., Ref.~\cite{nav00:62}.
\subsection{Hamiltonian and effective interactions}
The goal is to solve the $A$-body Schr\"odinger equation with an
intrinsic Hamiltonian of the form
\begin{equation}
    H_A= 
    \frac{1}{A}\sum_{i<j}^{A}\frac{(\vec{p}_i-\vec{p}_j)^2}{2m}
    + \sum_{i<j}^A V_{\nn, ij}  \; ,
    \label{eq:H}
\end{equation}
where $m$ is the nucleon mass and $V_{\nn, ij}$ is the \nn\ interaction
including both strong and electromagnetic components. As mentioned
earlier, we will not use three-body forces in this study since we strive
to maximize the size of the model space. By adding a center-of-mass (CM)
HO Hamiltonian $H_\mathrm{CM}^\Omega = T_\mathrm{CM} +
U_\mathrm{CM}^\Omega$ (where $U_\mathrm{CM}^\Omega = A m \Omega^2
\vec{R}^2 / 2, \; \vec{R} = \sum_{i=1}^A \vec{r} / A$), we facilitate
the use of a convenient HO basis. The modified Hamiltonian can be
separated into one- and two-body terms
%
%
\begin{equation}
  \begin{split}
    H_A^\Omega = & H_A + H_\mathrm{CM}^\Omega = \sum_{i=1}^A h_i +
    \sum_{i<j}^A V_{ij}^{\Omega,A} 
    \\ 
    = &\sum_{i=1}^A \left[ \frac{\vec{p}_i^2}{2m}
    +\frac{1}{2}m\Omega^2 \vec{r}^2_i
    \right] 
    \\
    &+ \sum_{i<j}^A \left[ V_{{\nn}, ij}
    -\frac{m\Omega^2}{2A} (\vec{r}_i-\vec{r}_j)^2 \right] \; .
    \label{eq:Homega}
  \end{split}
\end{equation}
The next step is to divide the $A$-nucleon infinite HO basis space into
an active, finite model space ($P$) and an excluded space ($Q = 1-
P$). The model space consists of all configurations with $\leq \nm \ho$
excitations above the unperturbed ground state. In this approach there
is no closed-shell core; meaning that all nucleons are active.

Since we solve the many-body problem in a finite model space, the
realistic \nn\ interaction will yield pathological results because of
the short-range repulsion. Consequently, we employ effective interaction
theory. In the \emph{ab initio} NCSM approach, the model-space dependent
effective interaction is constructed by performing a unitary
transformation of the Hamiltonian~\eqref{eq:Homega}, $e^{-S} H_A^\Omega
e^S$, such that the model space and the excluded space are decoupled $Q
e^{-S} H_A^\Omega e^S P = 0$. This procedure has been discussed by Lee
and Suzuki~\cite{suz80:64,suz82:68}, and yields a Hermitian effective
interaction $H_\mathrm{eff} = Pe^{-S} H_A^\Omega e^S P$ which acts in
the model space and reproduces exactly a subset of the eigenspectrum to
the full-space Hamiltonian. In general, this effective Hamiltonian will
be an $A$-body operator which is essentially as difficult to construct
as to solve the full $A$-body problem. In this study we approximate the
effective Hamiltonian at the two-body cluster level. The basic idea is
to derive it from high-precision solutions to the two-body problem with
$\mathcal{H}_2 = h_1 + h_2 + V_{12}^{\Omega,A}$, where the two-body term
is the same as in Eq.~\eqref{eq:Homega}. The final result will be a
two-body effective interaction $V_{12,\mathrm{eff}}^{\Omega,A}$. See
Ref.~\cite{nav00:62,cau02:66} for details.

We note that our approximated effective interaction will depend on the
nucleon number $A$, the HO frequency $\Omega$, and the size of the model
space which is defined by \nm. However, by construction, it will
approach the starting bare interaction $V_{ij,\mathrm{eff}}^{\Omega,A}
\to V_{ij}^{\Omega,A}$, as $\nm \to \infty$. Consequently, the dependence on
$\Omega$ will decrease with increasing model space, and the NCSM results
will converge to the exact solution. A very important feature of the
NCSM is the fact that the effective interaction is translationally
invariant so that the solutions can be factorized into a CM component
times a wave function corresponding to the internal degrees of
freedom. Due to this property it is straightforward to remove CM effects
completely from all observables.
\subsection{Solution of the many-body {S}chr\"odinger equation}
Once the effective interaction has been derived, we can construct the
effective Hamiltonian in the $A$-body space. In this process we
subtract the CM Hamiltonian $H_\mathrm{CM}^\Omega$, and add
the Lawson projection term $\beta(H_\mathrm{CM} - \frac{3}{2}\ho )$ to
shift eigenstates with excited CM motion up to high energies. States
with $0S$ CM motion are not affected by this term and, consequently,
their eigenenergies will be independent of the particular choice of
$\beta$. We are now left with a Hamiltonian of the form
%
%
\begin{multline}
  H_{A,\mathrm{eff}}^\Omega = P \Bigg\{ \sum_{i<j}^A 
  \bigg[ \frac{(\vec{p}_i-\vec{p}_j)^2}{2Am}
  +\frac{m\Omega^2}{2A} (\vec{r}_i-\vec{r}_j)^2 
  \\
  + V_{ij,\mathrm{eff}}^{\Omega,A} \bigg]  + 
  \beta \left( H_\mathrm{CM} - \frac{3}{2}\ho \right) \Bigg\} P \; .
  \label{eq:Homegaeff}
\end{multline}
%

The computational problem of obtaining the many-body eigenvalues is
non-trivial due to the very large matrix dimensions involved. The
largest model space that we encountered in this study was the
\nuc{11}{B} 9\ho\ (positive parity) space, for which the dimension
exceeds $d_P = 1.1 \times 10^9$. For \beel(\beni), the 9\ho\ space gives
$d_P = 7.1 \times 10^8(2.0 \times 10^8)$. To solve this problem we have
used a specialized version of the shell model code
\textsc{antoine}~\cite{cau99:30,cau99:59}, recently adapted to the
NCSM~\cite{cau01:64}. This code works in the $M$ scheme for basis
states, and uses the Lanczos algorithm for diagonalization. The number
of iterations needed to converge the first eigenstates is significantly
reduced by the implementation of a sophisticated strategy for selecting
the pivot vectors. This feature of the code is absolutely crucial when
using it to perform calculations in very large model spaces.

Furthermore, the code takes advantage of the fact that the dimensions of
the neutron and proton spaces are small with respect to the full
dimension. Therefore, before the diagonalization, all the matrix
elements involving one- and two-body operators acting in a single
subspace (proton or neutron) are calculated and stored. As a
consequence, during the Lanczos procedure, all non-zero proton-proton
and neutron-neutron matrix elements can be generated with a simple
loop. Furthermore, the proton-neutron matrix elements are obtained with
three integer additions~\cite{cau99:30}. However, for no-core
calculations (in which all nucleons are active) the number of shells
and, consequently, the number of matrix elements that are precalculated,
becomes very large. Consider, e.g., the \nuc{11}{B} calculation in the
9\ho\ space. The full dimension is $d_P = 1.1 \times 10^9$, while the
number of active shells is 66 and the total number of neutron plus
proton Slater determinants is $N(n) + N(p) = 1.0 \times 10^7$. This
corresponds to 80~Gb of precalculated and stored matrix elements. In
contrast, consider a shell-model calculation of \nuc{57}{Ni} in the full
$fp$ space. The total dimension is larger, $d_P = 1.4 \times 10^9$,
but there are only four active shells which gives $N(n) + N(p) = 1.8
\times 10^6$ and requires merely 1~Gb of precalculated data.

A recent development of the NCSM is the ability to further process the
wave functions, resulting from the shell-model calculation, to obtain
translationally invariant densities~\cite{nav04:70} and cluster form
factors~\cite{nav04:70_2}. The latter can be used to obtain
spectroscopic factors, but can also serve as a starting point for an
\emph{ab initio} description of low-energy nuclear reactions. We have
employed these new capabilities to study the overlap of \beel\ with
different $\nuc{10}{Be}+n$ channels.
\subsection{\label{sec:int}Realistic \nn\ interactions}
Four different, high-precision \nn\ interactions have been used in this
study. These are: the Argonne V8$^\prime$
(\avp)~\cite{wir95:51,pud97:56}, the CD-Bonn 2000
(\cdb)~\cite{mac01:63}, the N$^3$LO~\cite{ent03:68}, and the
INOY~\cite{dol04:69,dol03:67} potentials. We can divide these
interactions into three different types:

\emph{1. Local in coordinate space:} The \avp\ interaction is
an isospin-invariant subset of the phenomenological Argonne $v_{18}$
potential~\cite{wir95:51} plus a screened Coulomb potential. This
interaction is local in coordinate space and it is also employed in
the Green's Function Monte Carlo (GFMC)
approach~\cite{pud97:56}. Consequently, the use of this potential
allows for a direct comparison of results from the NCSM and the GFMC
methods.

\emph{2. Non-Local in momentum space:} The \cdb\
interaction~\cite{mac01:63} is a charge-dependent \nn\ interaction based
on one-boson exchange. It is described in terms of covariant Feynman
amplitudes, which are non-local. Consequently, the off-shell behavior of
the CD-Bonn interaction differs from commonly used local potentials
which leads to larger binding energies in nuclear few-body systems. The
newly developed \nlo\ interaction~\cite{ent03:68} is also represented by
a finite number of Feynman diagrams. This interaction, however, is based
on chiral perturbation theory at the fourth order, which means that it
is derived from a Lagrangian that is consistent with the symmetries of
QCD. It represents a novel development in the theory of nuclear
forces. It is particularly interesting to note that many-body forces
appear naturally already at the next-to-next-to-leading order (NNLO) of
this low-energy expansion. However, in this study we use solely the \nn\
part of the \nlo\ interaction. This \nn\ potential has previously been
applied in the NCSM approach to study the $p$-shell nuclei \nuc{6}{Li}
and \nuc{10}{B}~\cite{nav04:69}.

\emph{3. Non-Local in coordinate space:} A new type of interaction,
which respects the local behavior of traditional \nn\ interactions at
longer ranges but exhibits a non-locality at shorter distances, was
recently proposed by Doleschall \emph{et al}~\cite{dol04:69,dol03:67}.
The authors are exploring the extent to which effects of multi-nucleon
forces can be absorbed by non-local terms in the \nn\ interaction. Their
goal was to investigate if it is possible to introduce non-locality in
the \nn\ interaction so that it correctly describes the three-nucleon
bound states \nuc{3}{H} and \nuc{3}{He}, while still reproducing \nn\
scattering data with high precision. Note that all other \nn\
interactions give a large underbinding of $A \geq 3$ systems. In
practice, the \inoy\ interaction was constructed by combining an inner
($<3$~fm) phenomenological non-local part with a local Yukawa
tail. Hence the name INOY (Inside Nonlocal Outside Yukawa). The so
called IS version of this interaction, introduced in
Ref.~\cite{dol03:67}, contains short-range non-local potentials in
$^1S_0$ and $^3S_1-^{3\!\!}D_1$ partial waves while higher partial waves are
taken from Argonne $v_{18}$. In this study we are using the IS-M
version, which includes non-local potentials also in the $P$ and $D$
waves~\cite{dol04:69}. It is important to note that, for this particular
version, the on-shell properties of the triplet $P$-wave interactions
have been modified in order to improve the description of $3N$ analyzing
powers. The $^{3\!}P_0$ interaction was adjusted to become less attractive,
the $^3P_1$ became more repulsive, and the $^{3\!}P_2$ more
attractive. Unfortunately, this gives a slightly worse fit to the
Nijmegen $^{3\!}P$ phase shifts.
%
\section{\label{sec:results}Results}
By construction, the \emph{ab initio} NCSM method is guaranteed to
converge either by calculating the effective interaction using a fixed
cluster approximation (e.g., two-body) and then solving the eigenvalue
problem in increasing model spaces ($\nm \to \infty$), or by working in
a limited model space but increasing the clustering of the effective
interaction towards the full $A$-body one. Our codes are currently
constructed to derive effective interactions up to the level of
three-body clustering (with or without three-body forces). However, in
this study we have chosen to approach convergence by trying to maximize
our model space and, therefore, we limit ourselves to the use of
two-body effective interactions. Thus we are able to reach the 9\ho\
model space for nuclei with $A=11$. This maximal space corresponds to
basis dimensions of $d_P =$ $2.0 \times 10^8$ (\beni), $7.0 \times 10^8$
(\beel), and $1.1 \times 10^9$ (\nuc{11}{B}). For \nuc{13}{C}, which is
briefly discussed in connection to the parity-inversion problem, the
largest space that we were able to reach was $8\ho$ ($d_P = 8.2 \times
10^8$).

Note that model spaces with an even(odd) number of HO excitations give
negative-(positive-)parity states for the nuclei under study. When
constructing a full spectrum we combine the $\nm\ho$ and $(\nm+1)\ho$
results, with \nm\ being an even number. In connection to this, we should
also point out that very few states in \beni\ and \beel\ are particle
stable. However, in the NCSM approach, all states are artificially bound
due to the truncated model space and the use of HO basis functions.

In addition to a careful study of the level ordering in \beni\ and
\beel, with a particular focus on the position of the positive-parity
states, we also calculate electromagnetic moments and transition
strengths. For this we use traditional one-body transition operators
with free-nucleon charges. Note that, due to the factorization of our
wave function into CM and intrinsic components, we obtain
translationally invariant matrix elements for all observables that we
investigate in this work. However, we have not renormalized the
operators, which means that the results are not corrected for the fact
that we work in a truncated model space. The theoretical framework for
performing this renormalization is in place, and the process is
underway~\cite{ste04:nucl-th}. Until we are ready to implement the use
of effective operators, we can get an indication on the need for
renormalization by studying the basis-size dependence of our calculated
observables .

The cluster decomposition of the \beel\ ground state into
$\nuc{10}{Be}+n$ is of particular interest due to the small neutron
separation energy. We have employed the formalism recently developed in
Refs.~\cite{nav04:70,nav04:70_2} to calculate cluster overlap functions
using our NCSM wave functions. These results are presented in
Sec.~\ref{sec:11be}.
%
\subsection{\label{sec:hodep}Dependence on HO frequency}
\begin{figure*}[hbtp]
  \begin{minipage}{0.95\textwidth}
    \begin{minipage}[t]{0.47\textwidth}
      \centering
      \mbox{}\\
      \includegraphics*[width=\textwidth]{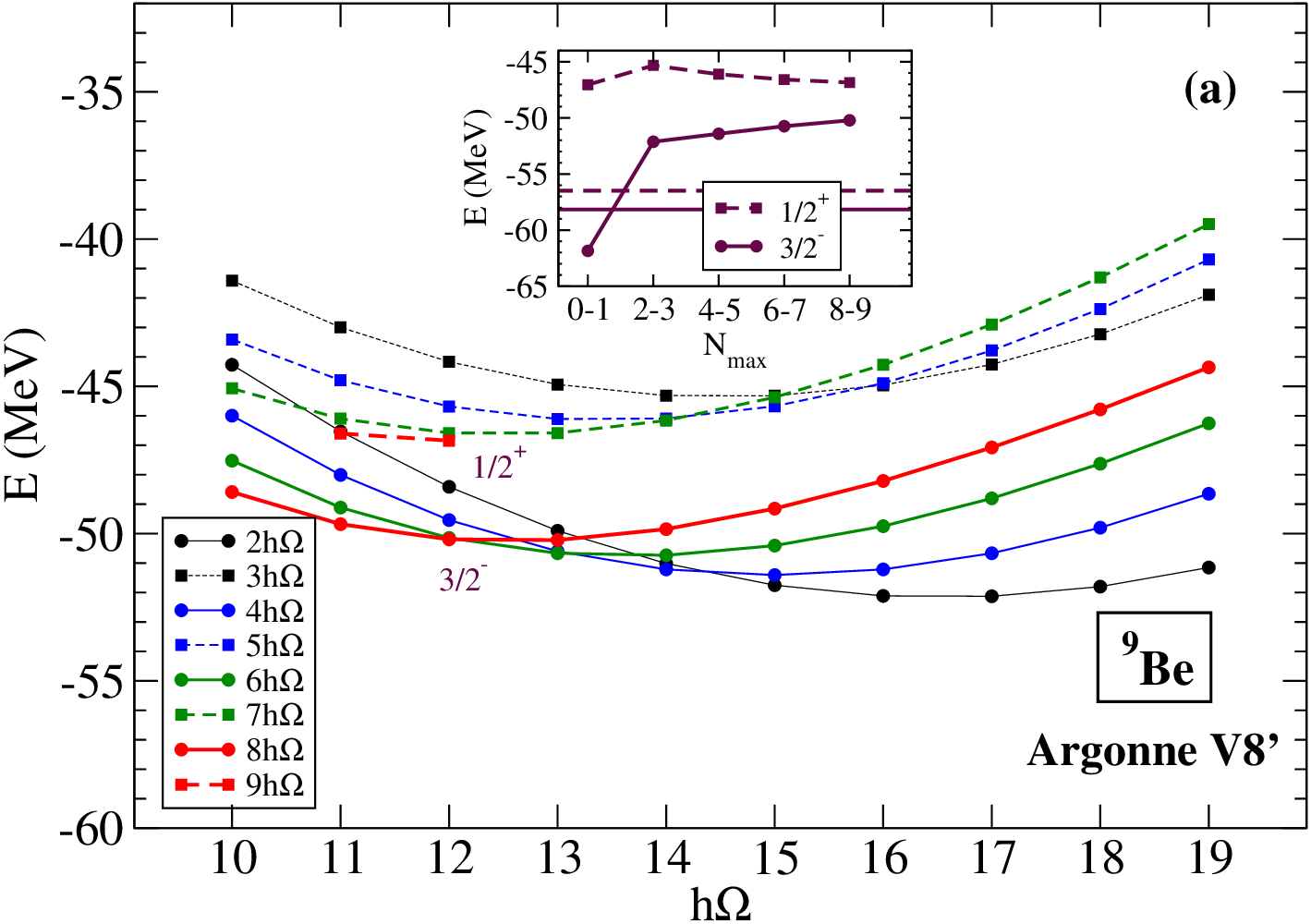}\\
      \includegraphics*[width=\textwidth]{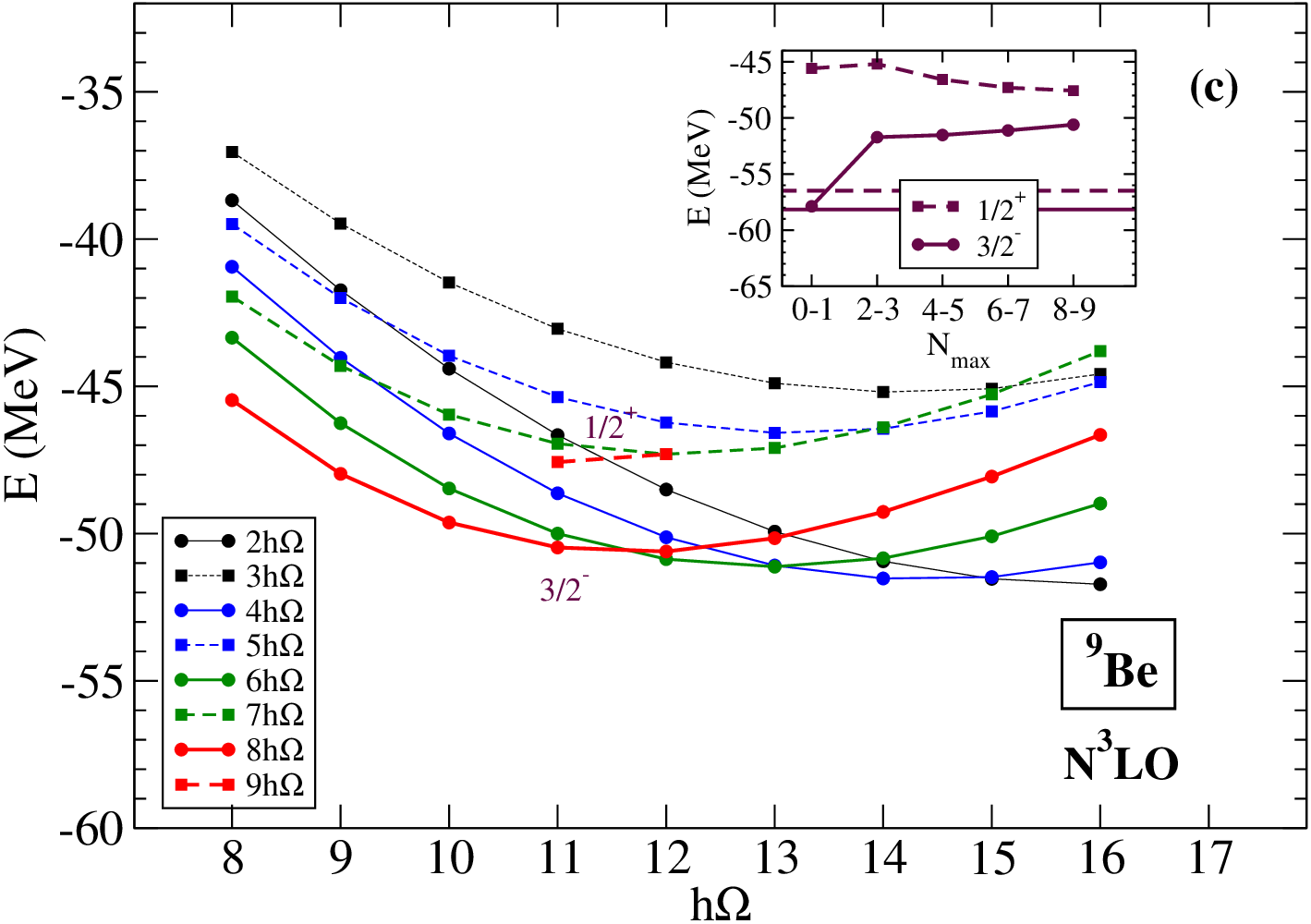}\\ 
      \mbox{}
    \end{minipage}
  \hfill
    \begin{minipage}[t]{0.47\textwidth}
      \centering
      \mbox{}\\
      \includegraphics*[width=\textwidth]{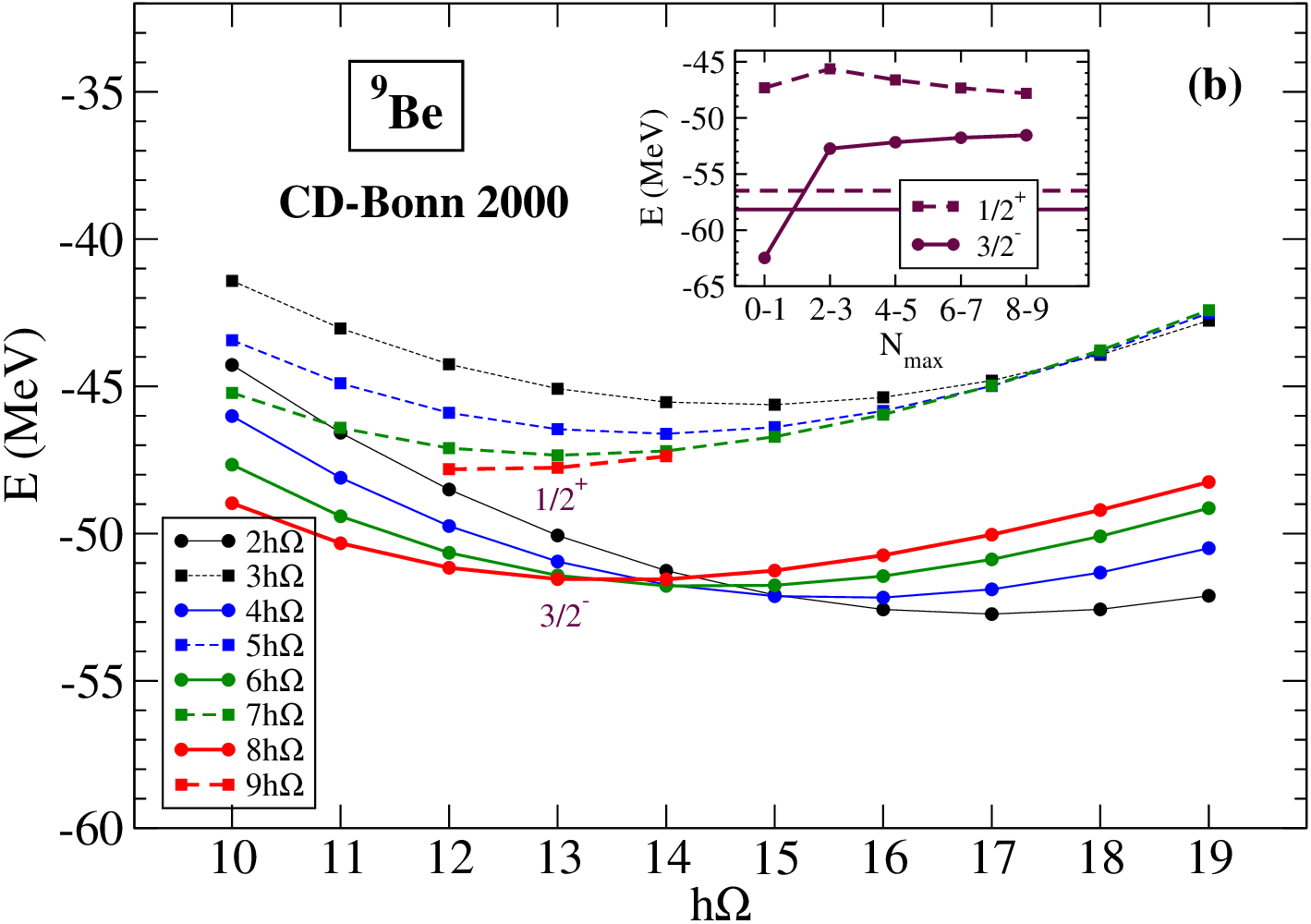}\\
      \includegraphics*[width=\textwidth]{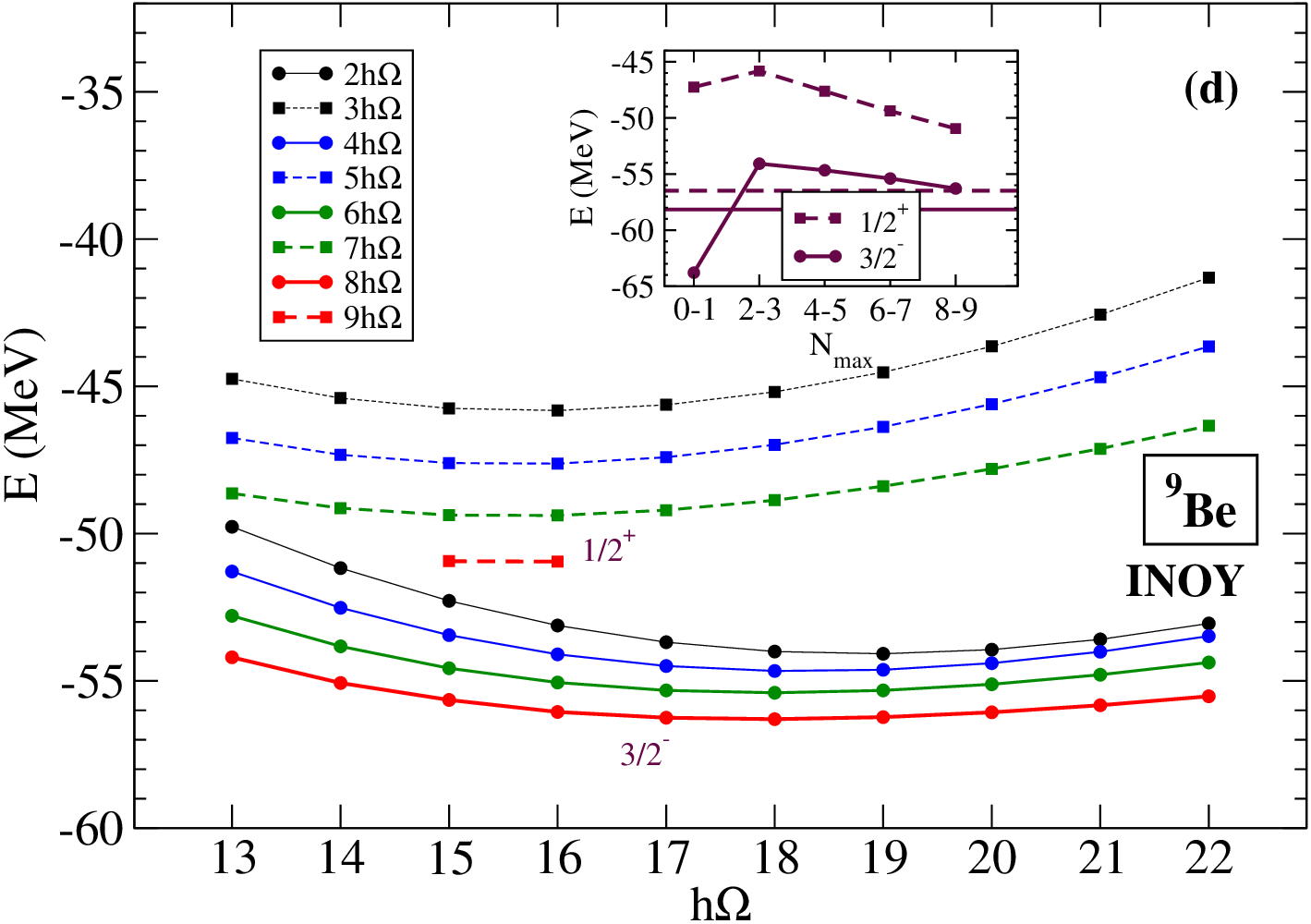}\\
      \mbox{}
    \end{minipage}
  \end{minipage}
  \caption{\co\ The dependence on HO frequency for the calculated
  \beni\jpt{3}{1}{-}{1} (solid lines) and \beni\jpt{1}{1}{+}{1} (dashed
  lines) binding energies. Each panel correspond to a particular \nn\
  interaction: (a) \avp, (b) \cdb, (c) \nlo, (d) \inoy; and each
  separate line corresponds to a specific model space. The insets
  demonstrate how the minima of the curves converge as the model space
  is increased. The horizontal lines are the experimental values.%
  \label{fig:9freqdep}}
\end{figure*}

\begin{figure*}[hbtp]
  \begin{minipage}{0.95\textwidth}
    \begin{minipage}[t]{0.47\textwidth}
      \centering
      \mbox{}\\
      \includegraphics*[width=\textwidth]{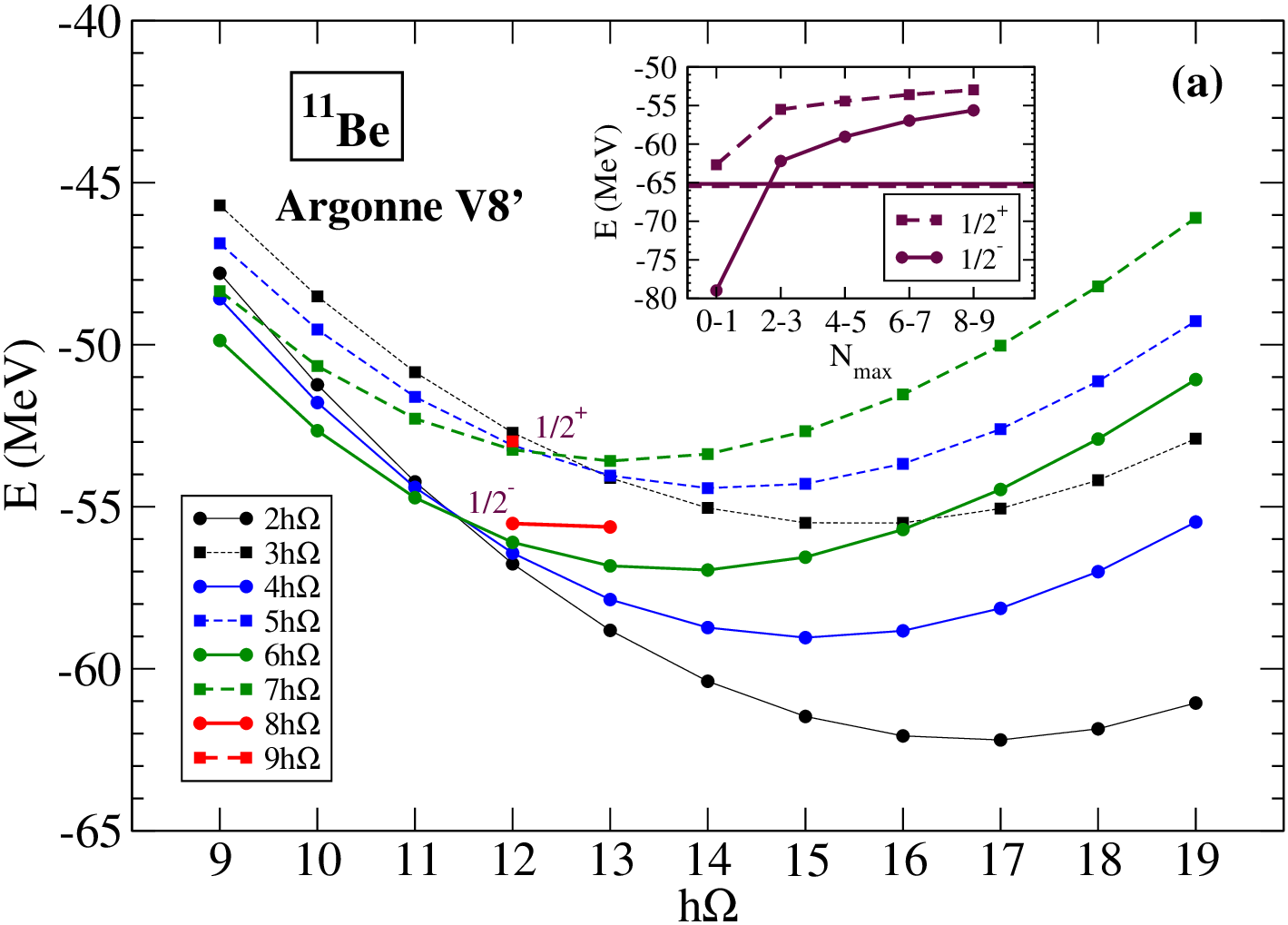}\\
      \includegraphics*[width=\textwidth]{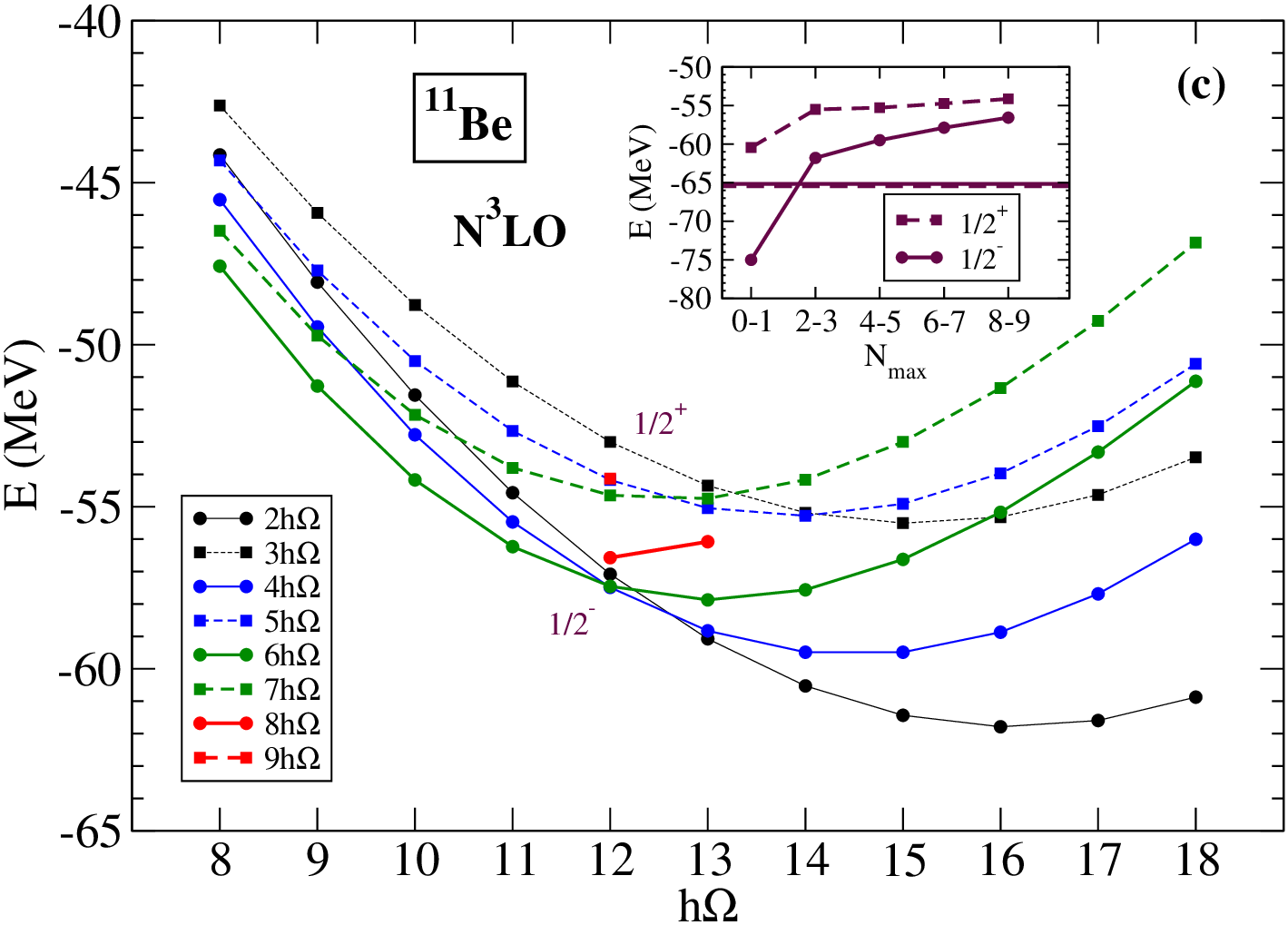}\\ 
      \mbox{}
    \end{minipage}
  \hfill
    \begin{minipage}[t]{0.47\textwidth}
      \centering
      \mbox{}\\
      \includegraphics*[width=\textwidth]{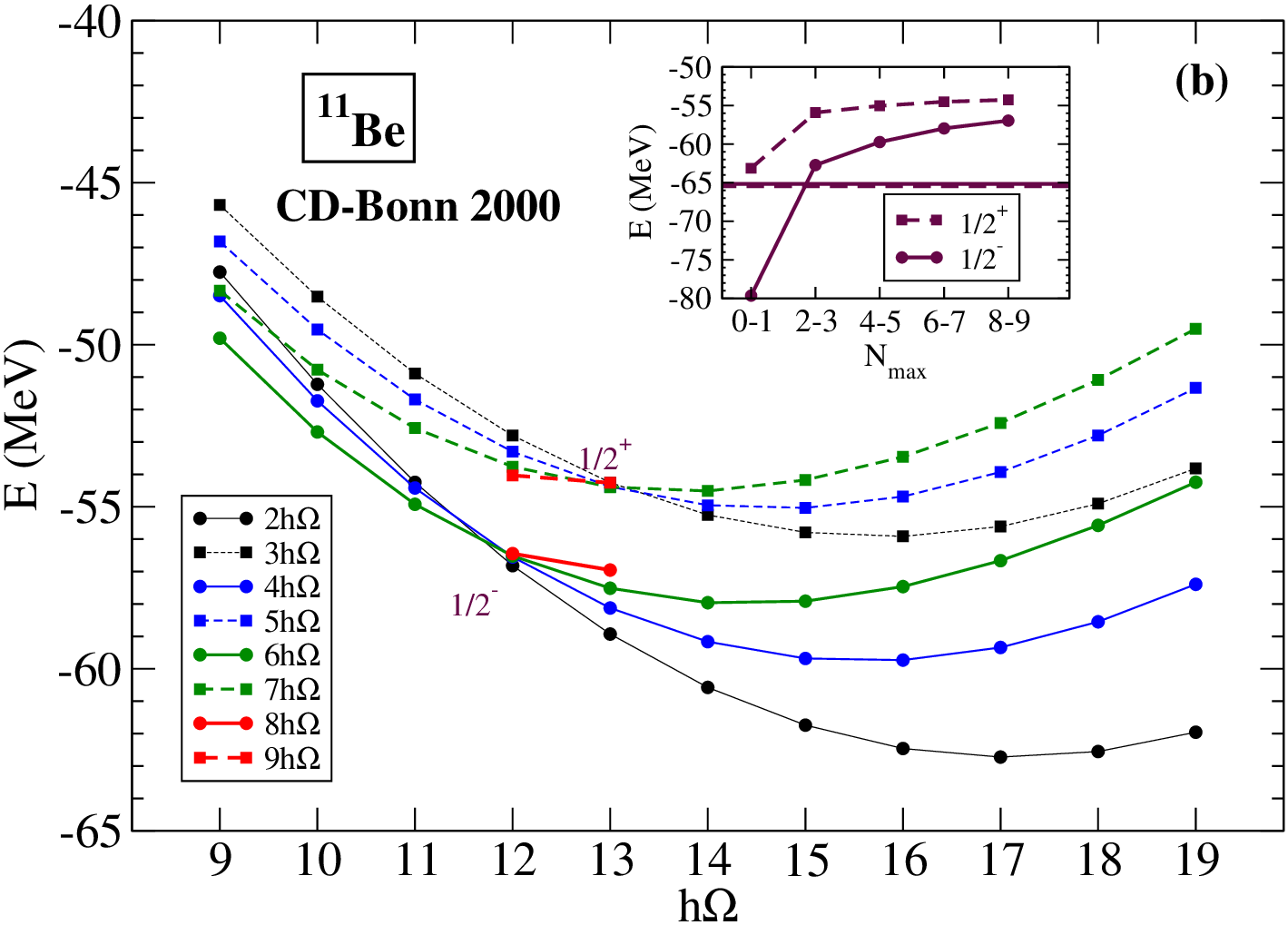}\\ 
      \includegraphics*[width=\textwidth]{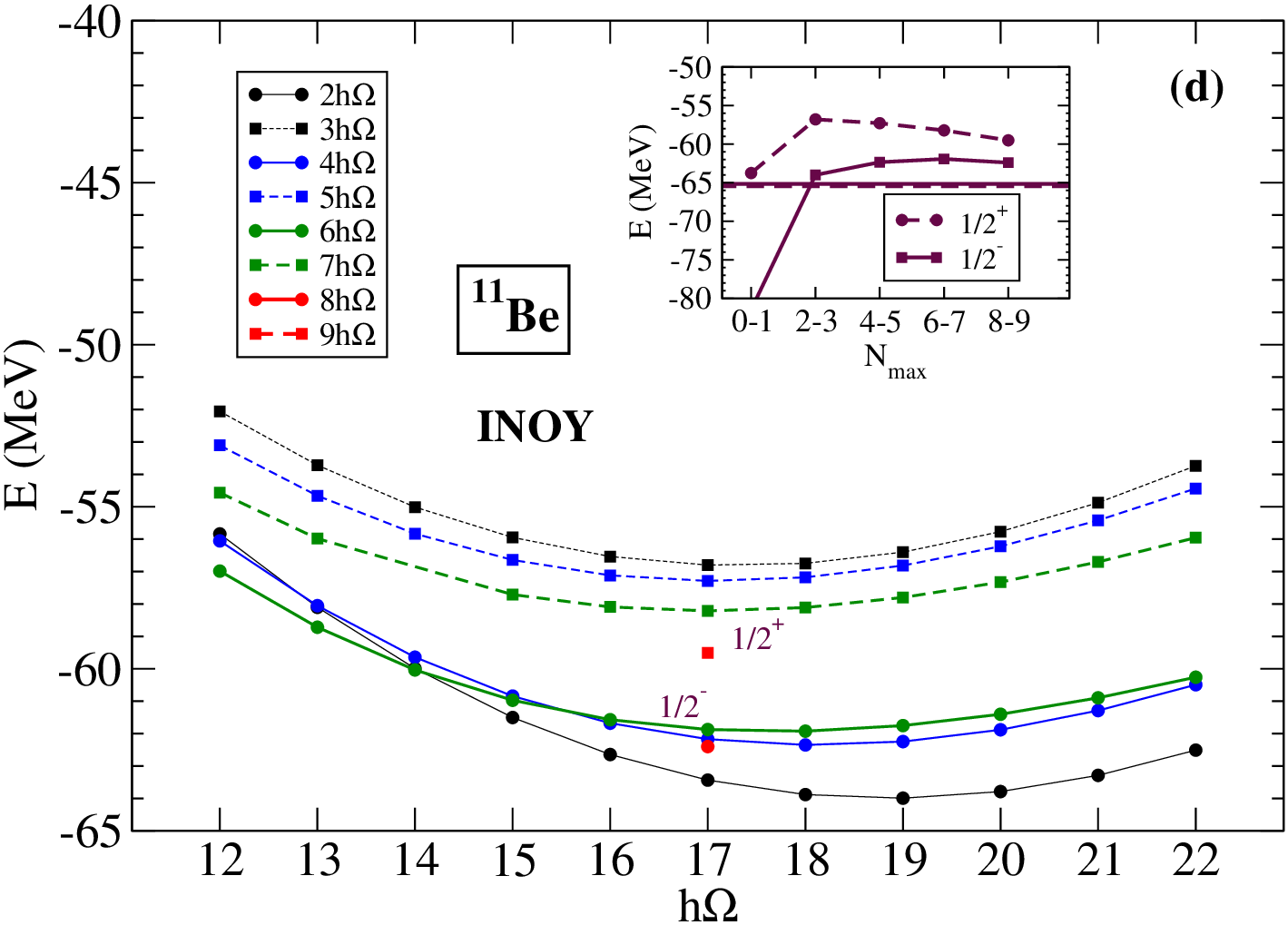}\\
      \mbox{}
    \end{minipage}
  \end{minipage}
  \caption{\co\ The dependence on HO frequency for the calculated
  \beel\jpt{1}{1}{-}{3} (solid lines) and \beel\jpt{1}{1}{+}{3} (dashed
  lines) binding energies. Each panel correspond to a particular \nn\
  interaction: (a) \avp, (b) \cdb, (c) \nlo, (d) \inoy; and each
  separate line corresponds to a specific model space. The insets
  demonstrate how the minima of the curves converge as the model space
  is increased. The horizontal lines are the experimental values.%
  \label{fig:11freqdep}}
\end{figure*}

%
The first step in our study is a search for the optimal HO frequency. In
principle, the intrinsic properties of the nucleus should not depend on
the particular value of \ho\ since it only enters the
Hamiltonian~\eqref{eq:Homega} through a CM-dependent term. In practice
however, due to the cluster approximation of the effective interaction,
our results will be sensitive to the choice of \ho. Furthermore, by
construction, the effective interactions depend on the size of the model
space, \nm, and on the number of nucleons, $A$. In order to investigate
these dependences we have performed a large series of calculations for a
sequence of frequencies. The results from this study are presented in
Figs.~\ref{fig:9freqdep} and \ref{fig:11freqdep} (for \beni\ and \beel,
respectively) as curves showing the frequency dependence of the binding
energy in different model spaces.  We have studied this dependence for
the lowest state of each parity.  We are looking for the region in which
the dependence on $\Omega$ is the smallest; and we select this frequency
(from the calculation in the largest model space) to use in our detailed
investigation of excited states. In our present case, this optimal
frequency always corresponds to an energy minimum. Note, however, that
the NCSM is not a variational method and the convergence of the binding
energy with increasing model space is not always from above.

Following this procedure for each nucleus and interaction, we obtain the
optimal HO frequencies that are listed in Table~\ref{tab:optfreq}.  A
few general remarks regarding the HO dependence, observed for the
different interactions in Figs.~\ref{fig:9freqdep} and
\ref{fig:11freqdep}, can be made: (1) Clear signs of convergence is
observed. The dependence on $\Omega$ becomes weaker with increasing size
of the model space, and the relative difference between the calculated
ground-state energies is in general decreasing. Furthermore, the optimal
frequencies for the largest model spaces of each parity (8\ho\ and 9\ho)
are approximately the same. This motivates our use of a single frequency
to compute both positive- and negative-parity states; (2) This single
frequency is found to be in the range of about \ho = 11--13~MeV for all
interactions except for \inoy, which seems to prefer a significantly
larger HO frequency (\ho = 16--17~MeV); (3) In general, the behavior of
the \avp, \cdb, and \nlo\ interactions are very similar, but with \nlo\
having the largest dependence on \ho; (4) As could be expected, since it
is the only \nn\ potential which is capable of reproducing $3N$ binding
energies, the \inoy\ interaction exhibits a distinctively different
behavior compared to the three others. The dependence on $\Omega$ is
encouragingly small, but the ground-state energy is still changing with
increasing basis size. This is particularly true for the positive-parity
state. We also note, from the insets of Figs.~\ref{fig:9freqdep} and
\ref{fig:11freqdep}, that it is the only interaction for which the
resulting binding energies are approaching the experimental values.
\begin{table}[hbtp]
  \caption{Selected optimal HO frequencies (in [MeV]). These choices are
    based on the frequency variation studies presented in
    Figs.~\ref{fig:9freqdep} and \ref{fig:11freqdep}.\vspace*{1ex}%
  \label{tab:optfreq}}
  \begin{ruledtabular}
    \begin{tabular}{ccccc}
      Nucleus &  \multicolumn{4}{c}{Interaction}                      \\   
                       & \inoy\    &  \cdb\    & \nlo\     & \avp\     \\
      \hline
      \beni            & 16        & 12        & 11        & 12        \\
      \beel            & 17        & 13        & 12        & 12        \\
    \end{tabular}
  \end{ruledtabular}
\end{table}
%
%
\subsection{\label{sec:9be}\beni}
%
By studying the HO frequency dependence of the \beni\ binding energy
obtained with different \nn\ interactions (see Fig.~\ref{fig:9freqdep})
it is clear that the \cdb\ results have a slightly better convergence
rate and a weaker HO frequency dependence than \avp\ and \nlo. The
\inoy\ results display an even weaker frequency dependence, but the
binding energy is still moving with increasing \nm. It is clear from
Table~\ref{tab:9energies} that all interactions, with the possible
exception of \inoy, underbind the system. Actually, by studying the
convergence rate of the \inoy\ results in Fig.~\ref{fig:9freqdep}d, it seems
as if this interaction will eventually overbind \beni. This observation
confirms that the additional binding, usually provided by $3N$ forces,
can be produced by the \inoy\ interaction. The other three \nn\
interactions underbind by 12--14\%. The local \avp\ potential was also
used in a recent GFMC study of negative-parity states in
\beni~\cite{pie02:66}, and we find an excellent agreement with their
ground-state binding energy (see Table~\ref{tab:9energies}).

\begin{figure}[hbtp]
  \includegraphics*[width=0.9\columnwidth]{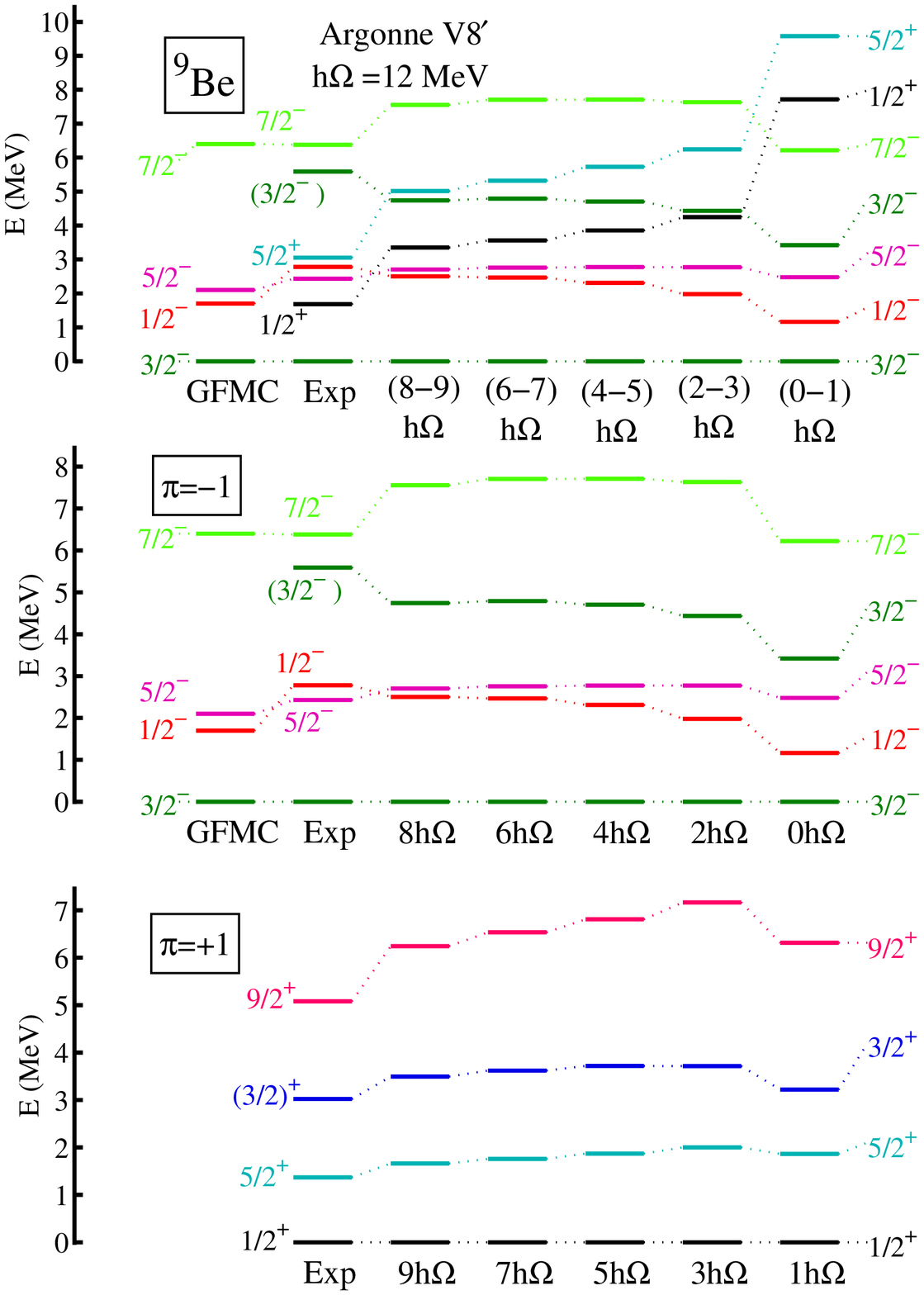}
  \caption{\co\ Excitation spectrum for \beni\ calculated using the
  \avp\ interaction in 0\ho--9\ho\ model spaces with a fixed HO
  frequency of $\hbar\Omega = 12$~MeV. The experimental values are from
  Ref.~\cite{til04:745}. The \avp\ results obtained by the GFMC
  method~\cite{pie02:66} are shown for comparison (note that only
  negative-parity states were computed). The two lower graphs show
  separately the negative- and positive-parity spectra, while the upper
  graph shows the combined spectrum with selected states.%
  \label{fig:9v8pspec}}
\end{figure}

\begin{figure}[hbtp]
  \includegraphics*[width=0.9\columnwidth]{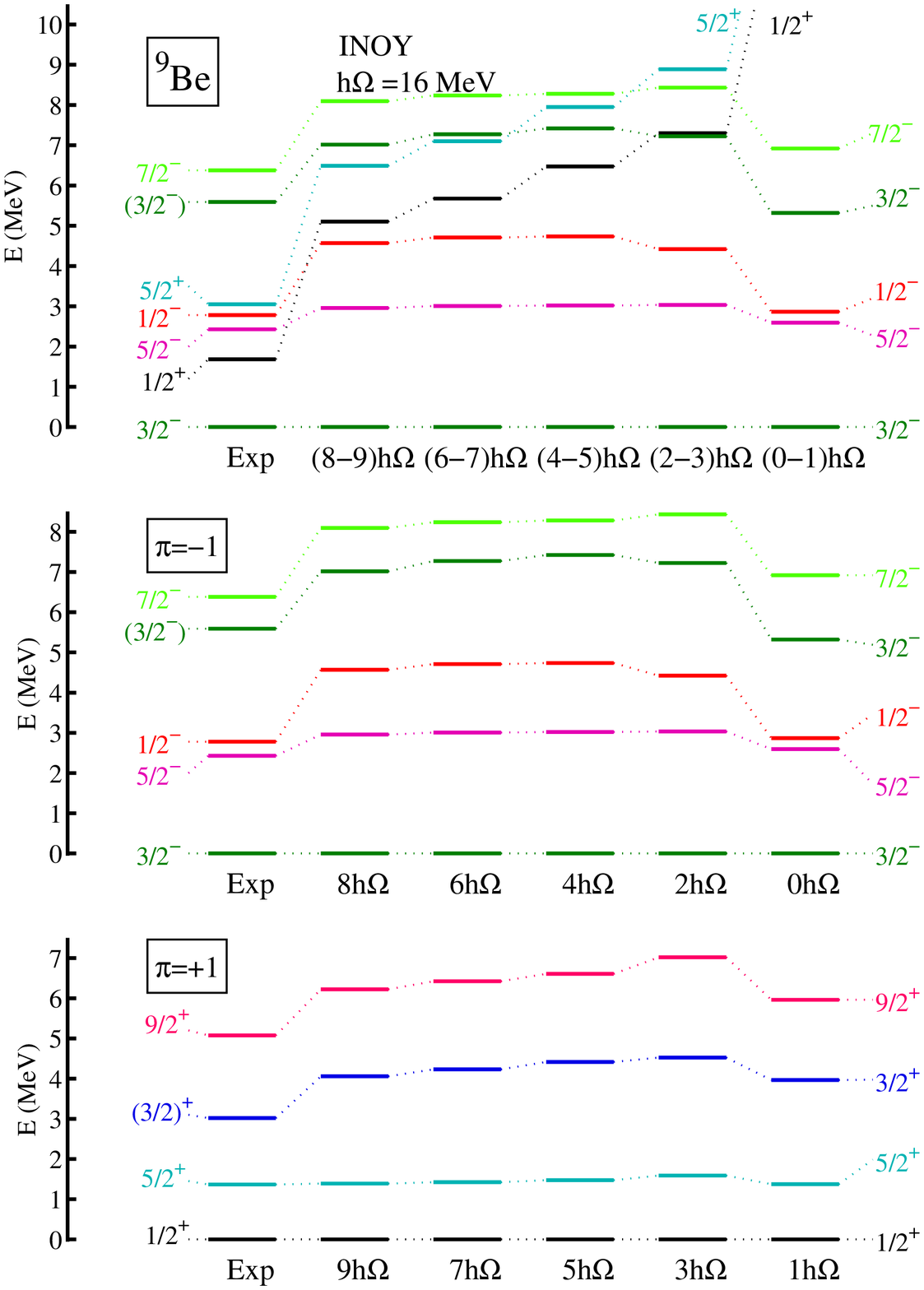}
  \caption{\co\ Excitation spectrum for \beni\ calculated using the
  \inoy\ interaction in 0\ho--9\ho\ model spaces with
  a fixed HO frequency of $\hbar\Omega = 16$~MeV. The experimental
  values are from Ref.~\cite{til04:745}. The two lower graphs show
  separately the negative- and positive-parity spectra, while the upper
  graph shows the combined spectrum with selected states.%
  \label{fig:9inoyspec}}
\end{figure}

In principle, the frequency dependence for each excited state should be
studied in order to compute its energy. This is particularly true in our
case where we want to compare negative- and positive-parity states. It
is therefore very encouraging that we find the same optimal frequency
for the first positive-parity state as for the negative-parity ground
state; and we select this frequency to use in our detailed investigation
of excited states. In Figs.~\ref{fig:9v8pspec},~\ref{fig:9inoyspec} and
Table~\ref{tab:9energies} we present our NCSM low-energy spectra for
different \nn\ interactions and compare the results to known
experimental levels.
\begin{table*}[hbtp]
  \caption{Experimental and calculated energies (in [MeV]) of the lowest
    negative- and positive-parity states in \beni. Quadrupole and
    magnetic moments (in [$e$fm$^2$] and [$\mu_N$]) for the ground
    state, as well as E2 and M1 strengths for selected transitions (in
    [$e^2$fm$^4$] and [$\mu_N^2$]). Results for the \avp, \cdb, \nlo\,
    and \inoy\ \nn\ interactions are presented. These calculations were
    performed in the 8(9)$\ho$ model space for
    negative-(positive-)parity states, using the HO frequencies listed
    in Table~\ref{tab:optfreq}. The GFMC results~\cite{pie02:66} are
    shown for comparison. Experimental values are
    from~\cite{til04:745}. $E_{x^+}$ denotes the excitation energy
    relative to the lowest positive-parity state.\vspace*{1ex}%
  \label{tab:9energies}}
  \begin{ruledtabular}
    \begin{tabular}{c@{\extracolsep\fill}cc@{\extracolsep{8ex}}ccc@{\extracolsep\fill}c}
      \beni\  &        & \multicolumn{4}{c}{NCSM}                      & GFMC  \\   
              & Exp    & \inoy\    &  \cdb\    & \nlo\     & \avp\     & \avp\ \\
      \hline
      $E_\mathrm{gs}\jpt{3}{1}{-}{1}$ 
              & -58.16 & -56.05    & -51.16    & -50.47    & -50.20    & -49.9(2) \\ 
      $E_x \jpt{3}{1}{-}{1}$
              & 0      & 0         & 0         & 0         & 0         & 0      \\
      $E_x \jpt{5}{1}{-}{1}$
              & 2.43   & 2.96      & 2.78      & 2.64      & 2.70      & 2.1      \\
      $E_x \jpt{1}{1}{-}{1}$
              & 2.78   & 4.57      & 2.68      & 2.33      & 2.50      & 1.7      \\
      $E_x \jpt{3}{2}{-}{1}$
              & 5.59\footnote[1]{The experimental spin-parity
      assignment of this level is ``less certain'' according to
      the TUNL Nuclear Data Evaluation~\cite{til04:745}.}   
                       & 7.02      & 4.98      & 4.53      & 4.74      & ---       \\
      $E_x \jpt{7}{1}{-}{1}$
              & 6.38   & 8.09      & 7.80      & 7.40      & 7.56      & 6.4      \\
      \hline
      $E\jpt{1}{1}{+}{1}$ 
              & -56.48 & -50.95    & -47.81    & -47.57    & -46.84    & ---       \\ 
      $E\jpt{1}{1}{+}{1} - E_\mathrm{gs}$
              & 1.68   & 5.10      & 3.35      & 2.90      & 3.35      & ---       \\
      $E_{x^+} \jpt{1}{1}{+}{1}$
              & 0    & 0       & 0       & 0       & 0       & ---       \\
      $E_{x^+} \jpt{5}{1}{+}{1}$
              & 1.37   & 1.39      & 1.68      & 1.68      & 1.66      & ---       \\
      $E_{x^+} \jpt{3}{1}{+}{1}$
              & 3.02\footnotemark[1]   
                       & 4.06      & 3.60      & 3.37      & 3.49      & ---       \\
      $E_{x^+} \jpt{9}{1}{+}{1}$
              & 5.08   & 6.22      & 6.36      & 6.21      & 6.24      & ---       \\
      \hline
      $Q_\mathrm{gs}$ 
              & 5.288(38)& 3.52      & 4.01      & 4.21      & 4.01      & 5.0(3) \\
      $\mu_\mathrm{gs}$ 
              & -1.1778(9)& -1.06    & -1.22     & -1.24     & -1.22     & -1.35(2) \\
      \hline
      $B \left( \mathrm{E}2; \frac{5}{2}_{_1}^- \to \frac{3}{2}_{_1}^- \right)$ 
              & 27.1(2) & 10.9     &  14.9     & 16.7      & 15.0      & --- \\
      $B \left( \mathrm{M}1; \frac{5}{2}_{_1}^- \to \frac{3}{2}_{_1}^- \right)$ 
              & 0.54(6)  & 0.46    &  0.37     & 0.36      & 0.37      & --- \\
    \end{tabular}
  \end{ruledtabular}
\end{table*}
As can be seen, the \avp, \cdb\ and \nlo\
interactions give the same level ordering and very similar excitation
energies. It is noteworthy that all these high-precision \nn\
interactions perform equally when applied to the $A=9$ system. We let
the \avp\ spectrum shown in Fig.~\ref{fig:9v8pspec} be the graphical
representation of all of them. Using the \avp\ we can also make a
comparison to the recent GFMC calculation~\cite{pie02:66}. 


In general, we observe a very reasonable agreement with experimental
levels of natural parity, while the unnatural-parity states are
consistently high in excitation energy. For both parities, there is a
general trend of convergence with increasing model space. When plotting
the negative- and positive-parity spectra separately, it is evident that
the relative level spacings are almost independent on the model space,
so that the level ordering within each parity projection is remarkably
stable. It is clear, however, that the relative position of the
negative- versus positive-parity states is still not
converged. Furthermore, when studying the \avp\ convergence pattern in
the upper panel of Fig.~\ref{fig:9v8pspec}, it seems as if this
interaction will predict the positive-parity states at too high
excitation energies even when the calculations will be converged. This
finding is consistent with an overall trend observed in other NCSM
calculations, and it has been speculated whether a three-body force will
correct this behavior~\cite{cau02:66}. Although we are still not able to
apply a true three-body force in a large enough model space, we get some
indications from the performance of the \inoy\ interaction. In
Fig.~\ref{fig:9inoyspec} we see that, for this interaction, the
positive-parity states are even higher in small model spaces, but that
they are also dropping much faster with increasing \nm. This issue is
investigated in further detail in Sec.~\ref{sec:pospar} where we discuss
the important question of parity inversion, and the general trend of the
position of natural- versus unnatural-parity states.

In addition to an increase in binding energy, it has been found that the
level ordering for many nuclei can be sensitive to the presence of
multi-nucleon forces~\cite{nav02:88,nav03:68,pie02:66}. This sensitivity
is the largest for those states where the spin-orbit interaction
strength is known to play a role. For \beni\ we find that our
calculations with the \avp, \cdb\ and \nlo\ interactions predict the
first-excited negative-parity state to be a $1/2^-$, while experiments
show that it is a $5/2^-$ (Note, however, that the \cdb\ interaction
predicts these two states to be almost degenerate, and to exhibit a
convergence trend indicating a possible level crossing at larger model
spaces.). This level reversal was also found in the GFMC calculations
using \avp. The \inoy\ interaction, on the other hand, gives the correct
level ordering, but instead overpredicts the spin-orbit splitting. By
performing a calculation in a smaller model space using the \avp\ plus
the Tucson-Melbourne TM$^\prime$(99)~\cite{coo01:30} three-nucleon
interaction, we found a similar result as with \inoy.

Our discussion up to this point has been concentrated on the low-lying
levels in \beni. However, in response to the recent evaluation published
by the TUNL Nuclear Data Evaluation Project~\cite{til04:745}, we have
also decided to summarize our results for higher excited states. It is
important to note that the experimental widths of these states are
generally quite large, and to compute them correctly with the NCSM
method requires a very large model space. Furthermore, at high
excitation energies, it is very probable that there will be some
admixture of 2\ho\ intruders, and these are usually predicted too high
in the NCSM. In any case, our results can serve as an important
guideline as to which $p$-shell states that can be expected to appear in
the spectrum, and consequently should be looked for in experiments. In
Table~\ref{tab:9energiescdb}, we present all levels that we have
calculated using the \cdb\ interaction in the 8\ho\ and 9\ho\ model
spaces. We also show the tabulated experimental levels below $E_x =
13$~MeV, taken from the most recent evaluation~\cite{til04:745}. A quick
comparison with the previous, published evaluation~\cite{ajz88:490}
(from 1988), reveals that several new levels have been discovered and
many spin-parity assignments have been changed. Altogether, these
changes lead to a much better agreement with our results. In the
negative-parity spectrum, our calculations give the correct level
ordering for the first six states. In particular, we correctly reproduce
the second $3/2^-$ and $5/2^-$ states that were introduced in the new
evaluation. On the other hand, we find a third $3/2^-$ state and a
second $1/2^-$ state that have not been observed in
experiments. However, these states are not fully converged in our 8\ho\
calculation, and they are still moving towards higher excitation
energies. We also find a $9/2^-$ state, which is fairly stable, and that
has not been experimentally identified. In the positive-parity spectrum,
the 6.76~MeV level has now been changed to being a $9/2^+$ which agrees
well with our level ordering. Finally, it is interesting to note that
our second $1/2^+$, $3/2^+$, and $5/2^+$ levels all appear in between
the first $9/2^+$ and $7/2^+$. None of these three states have, however,
been experimentally confirmed.
\begin{table}[hbtp]
  \caption{Experimental and calculated energies (in [MeV]) of the lowest
    negative- and positive-parity states in \beni. The calculations were
    performed in the 8(9)$\ho$ model space for
    negative-(positive-)parity states, using the \cdb\ \nn\ interaction
    with $\ho=12$~MeV. This table represents a more complete compilation
    of our computed levels (albeit for only one interaction) as compared
    to Table~\ref{tab:9energies}. Experimental values are
    from~\cite{til04:745}. $E_{x^+}$ denotes the excitation energy
    relative to the lowest positive-parity state.\vspace*{1ex}%
  \label{tab:9energiescdb}}
  \begin{ruledtabular}
    \begin{tabular}{ccc@{\hspace\fill}ccc}
      \multicolumn{3}{c}{Negative parity states} & 
      \multicolumn{3}{c}{Positive parity states} \\
      \beni\          & Exp    &  \cdb     & \beni          & Exp    &  \cdb    \\
      \hline
      $E_\mathrm{gs}\jpt{3}{1}{-}{1}$ 
                      & -58.16 & -51.16    & $E\jpt{1}{1}{+}{1}$ 
                                                            & -56.48 & -47.81    \\ 
      $E_x \jpt{3}{1}{-}{1}$
                      & 0      & 0         & $E\jpt{1}{1}{+}{1} - E_\mathrm{gs}$
                                                            & 1.68   & 3.35      \\
      $E_x \jpt{5}{1}{-}{1}$
                      & 2.43   & 2.78      & $E_{x^+} \jpt{1}{1}{+}{1}$
                                                            & 0    & 0           \\
      $E_x \jpt{1}{1}{-}{1}$
                      & 2.78   & 2.68      & $E_{x^+} \jpt{5}{1}{+}{1}$
                                                            & 1.37   & 1.68      \\
      $E_x \jpt{3}{2}{-}{1}$
                      & 5.59\footnote[1]{The experimental spin-parity
      assignment of this level is ``less certain'' according to
      the TUNL Nuclear Data Evaluation~\cite{til04:745}.}   
                               & 4.98      & $E_{x^+} \jpt{3}{1}{+}{1}$
                                                 & 3.02\footnotemark[1]   & 3.60      \\
      $E_x \jpt{7}{1}{-}{1}$
                      & 6.38   & 7.80      & $E_{x^+} \jpt{9}{1}{+}{1}$
                                                            & 5.08   & 6.36      \\

      $E_x \jpt{5}{2}{-}{1}$
                      & 7.94\footnotemark[1]   & 7.96      & $E_{x^+} \jpt{5}{2}{+}{1}$
                                                            &        & 7.66\footnote[2]
		      {Calculated in a smaller, 7$\ho$, model space.}         \\
      $E_x \jpt{3}{3}{-}{1}$
                      &        & 11.26     & $E_{x^+} \jpt{3}{2}{+}{1}$
                                                            &        & 7.91\footnotemark[2]      \\
      $E_x \jpt{1}{2}{-}{1}$
                      &        & 11.86     & $E_{x^+} \jpt{1}{2}{+}{1}$
                                                            &        & 8.13\footnotemark[2]      \\
      $E_x \jpt{9}{1}{-}{1}$
                      &        & 12.45     & $E_{x^+} \jpt{7}{1}{+}{1}$
                                                            &        & 8.48      \\
      $E_x \jpt{7}{2}{-}{1}$
                      & 11.28\footnotemark[1]  & 12.61     \\
      $E_x \jpt{5}{3}{-}{1}$
                      & 11.81  & 13.02     \\
    \end{tabular}
  \end{ruledtabular}
\end{table}

In Table~\ref{tab:9energies}, we also present our results for the
ground-state quadrupole and magnetic moments, as well as for selected
electromagnetic transition strengths. We should stress that free-nucleon
charges have been used in these calculations, and that the operators
have not been renormalized. On the other hand, the stability of our
results can be judged by investigating the dependence on the model
space. We find that the calculated ground-state magnetic moment and the
$B(\mathrm{M1};\frac{5}{2}_{_1}^- \to \frac{3}{2}_{_1}^-)$ transition
strength are almost converged, and in fair agreement with the
experimental values. The results for electromagnetic quadrupole
observables are, however, steadily increasing with basis size
enlargement and should clearly benefit from the use of effective
operators. As an example we can consider the evolution of the \nlo\
results: For the $\left\{4\:-\:6\:-\:8\right\}\ho$ sequence of model
spaces these observables increase as $Q_\mathrm{gs} =
\left\{+3.96\:-\:+4.10\:-\:+4.21\right\}$~[$e$fm$^2$], and $B \left(
\mathrm{E}2; \frac{5}{2}_{_1}^- \to \frac{3}{2}_{_1}^- \right) =
\left\{14.9\:-\:15.7\:-\:16.7\right\}$~[$e^2$fm$^4$]. We would also like
to highlight the fact that \inoy\ gives much smaller values for
$Q_\mathrm{gs}$ and $B(\mathrm{E}2)$ than the other interactions. This
is partly due to the fact that, for \inoy, our selected frequency is
much larger than for the other potentials which, in our limited model
space, corresponds to a smaller radial scale. In principle, a HO
frequency dependence study should be made for each operator. However, we
have also applied the \inoy\ interaction in studies of $A=3,4$ systems,
for which convergence can be easily reached. It was found that, in
particular the rms proton radius is always underpredicted. The same
result was obtained in Ref.~\cite{laz04:70} through exact solutions of
the Faddeev-Yakubovski equations, and it demonstrates that the
interaction is too soft, resulting in a faster condensation of nuclear
matter. In this work, we have studied the \beni\ point-nucleon radii as
well as the strong E1 transition from the first-excited to the ground
state using the \avp\ interaction. However, since these results will be
compared to \beel\ data, we postpone the discussion to
Sec.~\ref{sec:11be}.

In Tables~\ref{tab:9conf} and~\ref{tab:9spocc}, we show the resulting
$N\ho$-configurations and the single-particle occupancies of the \beni\
wave function, obtained with the four different interactions. Although
these quantities are not physical observables, they can still give
interesting information. We see that the wave functions obtained with
the \avp, \cdb\ and \nlo\ interactions are almost identical, while the
\inoy\ wave function has a considerably larger fraction of low-\ho\
excitations. This fact is in part due to the higher HO frequency being
used in the \inoy\ calculations. Furthermore, from Table~\ref{tab:9conf}
we see that this particular interaction gives a different distribution
of \ho\ excitations for the \jpt{3}{1}{-}{1} and \jpt{1}{1}{+}{1}
states; which would indicate that the latter is slightly more
deformed. However, this behavior is not observed for the other
interactions. The differences in occupations of single-particle levels
reflect some properties of the interactions. The fact that the
$0p_{3/2}$ and $0d_{5/2}$ levels have a larger occupation in the \inoy\
eigenstates is direct evidence for a stronger spin-orbit interaction.
\begin{table}[hbtp]
  \caption{Calculated configurations of the first negative- and
    positive-parity states in \beni. Results obtained in our largest
    model spaces (8$\hbar\Omega$ and 9$\hbar\Omega$, respectively) are
    presented. The calculations were performed with the HO frequencies
    listed in Table~\ref{tab:optfreq}.\vspace*{1ex}%
  \label{tab:9conf}}
  \begin{ruledtabular}
    \begin{tabular}{lccccc}
      & \multicolumn{5}{c}{\beni\jpt{3}{1}{-}{1} (8$\hbar\Omega$ model space)}\\
      \nn\ interaction & 0$\hbar\Omega$ & 2$\hbar\Omega$ & 4$\hbar\Omega$
        & 6$\hbar\Omega$ & 8$\hbar\Omega$ \\
      \hline
      \inoy\ & 0.58 & 0.19 & 0.13 & 0.06 & 0.04 \\
      \cdb\  & 0.49 & 0.22 & 0.15 & 0.08 & 0.06 \\
      \nlo\  & 0.47 & 0.24 & 0.16 & 0.08 & 0.06 \\
      \avp\  & 0.49 & 0.22 & 0.15 & 0.08 & 0.06 \\[2ex]
      \hline
      & \multicolumn{5}{c}{\beni\jpt{1}{1}{+}{1} (9$\hbar\Omega$ model space)}\\
      \nn\ interaction & 1$\hbar\Omega$ & 3$\hbar\Omega$ & 5$\hbar\Omega$
        & 7$\hbar\Omega$ & 9$\hbar\Omega$ \\
      \hline
      \inoy\ & 0.53 & 0.22 & 0.14 & 0.07 & 0.04 \\
      \cdb\  & 0.48 & 0.23 & 0.15 & 0.08 & 0.06 \\
      \nlo\  & 0.46 & 0.24 & 0.16 & 0.08 & 0.05 \\
      \avp\  & 0.49 & 0.23 & 0.15 & 0.08 & 0.06 \\[2ex]
    \end{tabular}
  \end{ruledtabular}
\end{table}
\begin{table}[hbtp]
  \caption{Calculated occupations of neutron single-particle levels for
    the first negative- and positive-parity states in \beni. Results
    obtained in our largest model spaces (8$\hbar\Omega$ and
    9$\hbar\Omega$, respectively) are presented. The calculations were
    performed with the HO frequencies listed in
    Table~\ref{tab:optfreq}.\vspace*{1ex}%
  \label{tab:9spocc}}
  \begin{ruledtabular}
    \begin{tabular}{lcccccc}
      & \multicolumn{5}{c}{\beni\jpt{3}{1}{-}{1} (8$\hbar\Omega$ model space)}\\
      \nn\ interaction & $0s_{1/2}$ & $0p_{1/2}$ & $0p_{3/2}$
        & $1s_{1/2}$ & $0d_{3/2}$ & $0d_{5/2}$ \\
      \hline
      \inoy\ & 1.804& 0.454& 2.380& 0.043& 0.064& 0.069\\
      \cdb\  & 1.773& 0.511& 2.277& 0.067& 0.060& 0.072\\
      \nlo\  & 1.768& 0.521& 2.256& 0.077& 0.060& 0.072\\
      \avp\  & 1.778& 0.519& 2.273& 0.064& 0.061& 0.071\\[2ex]
      \hline
      & \multicolumn{5}{c}{\beni\jpt{1}{1}{+}{1} (9$\hbar\Omega$ model space)}\\
      \nn\ interaction & $0s_{1/2}$ & $0p_{1/2}$ & $0p_{3/2}$
        & $1s_{1/2}$ & $0d_{3/2}$ & $0d_{5/2}$ \\
      \hline
      \inoy\ & 1.792 & 0.498 & 1.428 & 0.573 &
      0.111 & 0.379 \\
      \cdb\  & 1.761& 0.530& 1.377& 0.666& 0.113& 0.317\\
      \nlo\  & 1.757& 0.536& 1.365& 0.676& 0.117& 0.312\\
      \avp\  & 1.765& 0.535& 1.376& 0.664& 0.113& 0.313 \\[2ex]
    \end{tabular}
  \end{ruledtabular}
\end{table}
\subsection{\label{sec:11be}\beel}
Most of the observations made for \beni\ concerning the frequency
dependence of the calculated binding energies also hold true for
\beel. In general, however, the sensitivity to \ho\ is stronger in the
\beel\ case. Another important remark, that can be made from studying
Fig.~\ref{fig:11freqdep}, is that the binding energy of the first
positive-parity state calculated with the \inoy\ interaction is clearly not
converged. The relative shift in energy is actually slowly increasing
with model-space enlargement.

The experimental ground state of \beel\ is an intruder $1/2^+$ level,
while the first $p$-shell state is a $1/2^-$ situated at
$E_x=320$~keV. The neutron separation energy is only 503~keV, and there
are no additional bound states. This level-ordering anomaly constitutes
the famous parity-inversion problem. A number of excited states have
been observed in different reactions and beta-decay studies. However, as
can be seen from the summary presented in Table~\ref{tab:explevels},
there are considerable ambiguities in the spin-parity assignments.
\begin{table*}[hbtp]
  \caption{Present situation of the spin-parity assignments for the
    lowest states in \beel. The table contains published results from
    the FAS evaluation of 1990~\cite{ajz90:506rev} and from more recent
    experimental studies. These studies include direct reactions such as
    $(t,p)$ (Liu-Fortune) and $\nuc{12}{C}(\beel,\beel')$ (Fukuda) in
    which the extracted angular distributions were analyzed using DWBA
    theory. The remaining references are measurements of $\beta$-delayed
    neutrons in coincidence with $\gamma$-rays. All decays that were
    observed in these experiments had $\log(ft)$ values that were
    consistent with allowed transitions, indicating that the
    corresponding final states have negative parity and $J \leq
    \frac{5}{2}$. \vspace*{1ex}%
  \label{tab:explevels}}
  \begin{ruledtabular}
    \begin{tabular}{lccccccccc}
         & \multicolumn{9}{c}{States (MeV)} \\
    Ref. & 0.0 & 0.32 & 1.78 & 2.69 & 3.41 & 3.89 & 3.96 & 5.24 & 5.86
    \\
    \hline
    Ajz.-Sel.~\cite{ajz90:506rev} & $\frac{1}{2}^+$ & $\frac{1}{2}^-$ &
    $\left( \frac{5}{2} , \frac{3}{2} \right)^+$ & $\left(
    \frac{1}{2},\frac{3}{2},\frac{5}{2}^+\right)$ & $\left(
    \frac{1}{2},\frac{3}{2},\frac{5}{2}^+\right)$ & $\geq \frac{7}{2}$ &
    $\frac{3}{2}^-$ &     &     \\
    Liu~\cite{liu90:42} & $\frac{1}{2}^+$ & $\frac{1}{2}^-$ & $\frac{5}{2}^+$ &
    $\frac{3}{2}^-$ & $\frac{3}{2}^-$ & $\frac{3}{2}^+$ &
    $\frac{3}{2}^-$ & $\frac{5}{2}^-$ &
    $\left(\frac{1}{2}^+,\frac{1}{2}^- \right)$ \\
    Morrisey~\cite{mor97:627} & $\frac{1}{2}^+$ & $\frac{1}{2}^-$ & $(+)$ & $(-)$ & $(+)$
    & $(-)$ & $(-)$ & $(-)$ & $(-)$ \\
    Aoi~\cite{aoi97:616} & $\frac{1}{2}^+$ & $\frac{1}{2}^-$ & $\frac{5}{2}^+$ &
    $\frac{3}{2}^-$ & $\frac{3}{2}^-$ & $\frac{3}{2}^+$ &
    $\frac{3}{2}^-$ & $\frac{5}{2}^-$ &     \\
    Hirayama~\cite{hir04:738} & $\frac{1}{2}^+$ & $\frac{1}{2}^-$ & $(+)$ &
    $\frac{3}{2}^-$ & $\frac{3}{2}^-$ & $\frac{5}{2}^-$ &
    $\frac{3}{2}^-$ & $\frac{5}{2}^-$ &     \\
    Fukuda~\cite{fuk04:70} & $\frac{1}{2}^+$ & $\frac{1}{2}^-$ & $\left(
      \frac{3}{2} , \frac{5}{2} \right)^+$ &     & $\left(
      \frac{3}{2} , \frac{5}{2} \right)^+$ &     &     &     &     \\
   \end{tabular}
  \end{ruledtabular}
\end{table*}
\begin{figure}[hbtp]
  \includegraphics*[width=0.9\columnwidth]{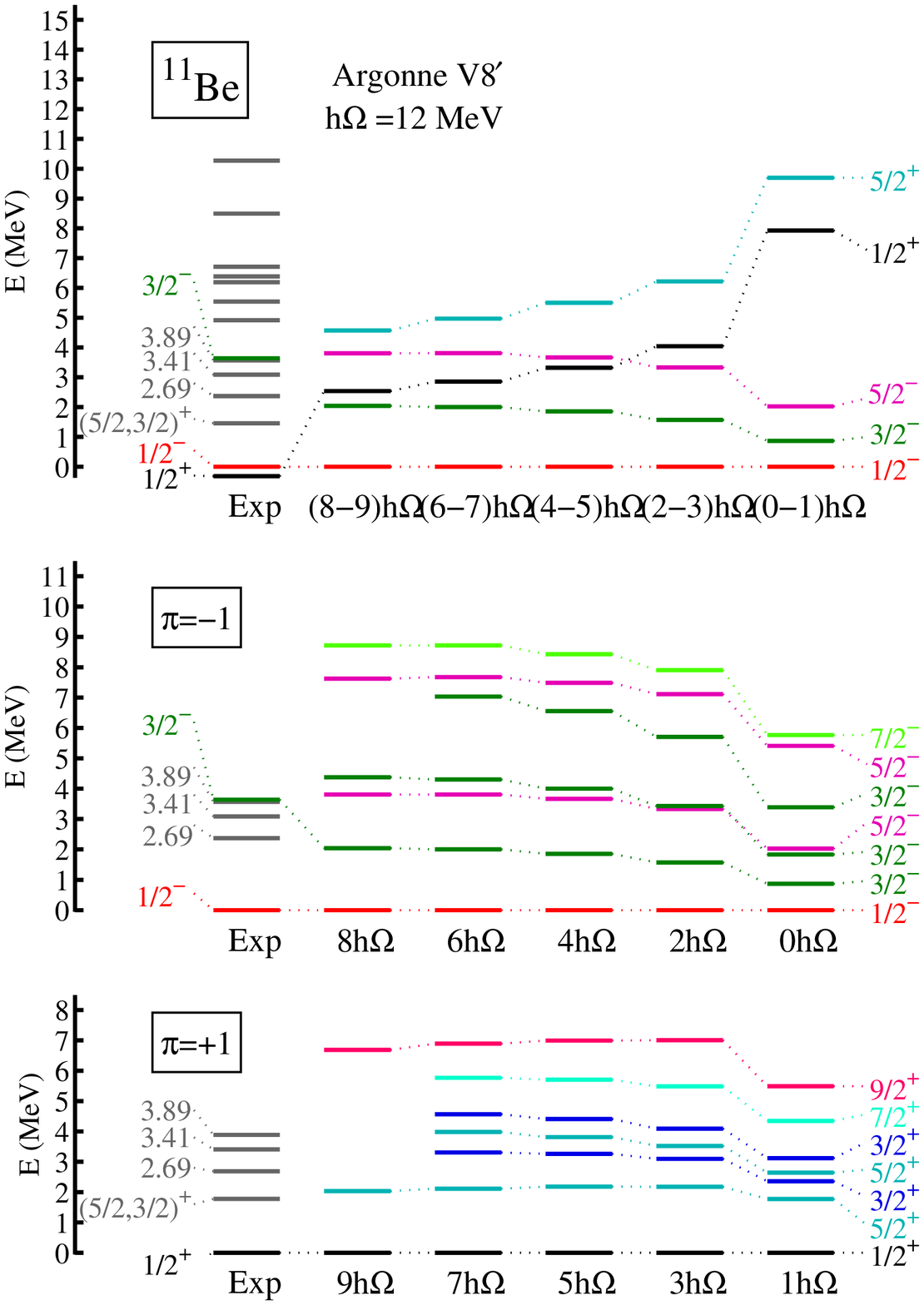}
  \caption{\co\ Excitation spectrum for \beel\ calculated using the
  \avp\ interaction in 0\ho--9\ho\ model spaces with
  a fixed HO frequency of $\hbar\Omega = 12$~MeV. The experimental
  values are from Ref.~\cite{ajz90:506rev}. The two lower graphs show
  separately the negative- and positive-parity spectra, while the upper
  graph shows the combined spectrum with selected states.%
  \label{fig:11v8pspec}}
\end{figure}
\begin{figure}[hbtp]
  \includegraphics*[width=0.9\columnwidth]{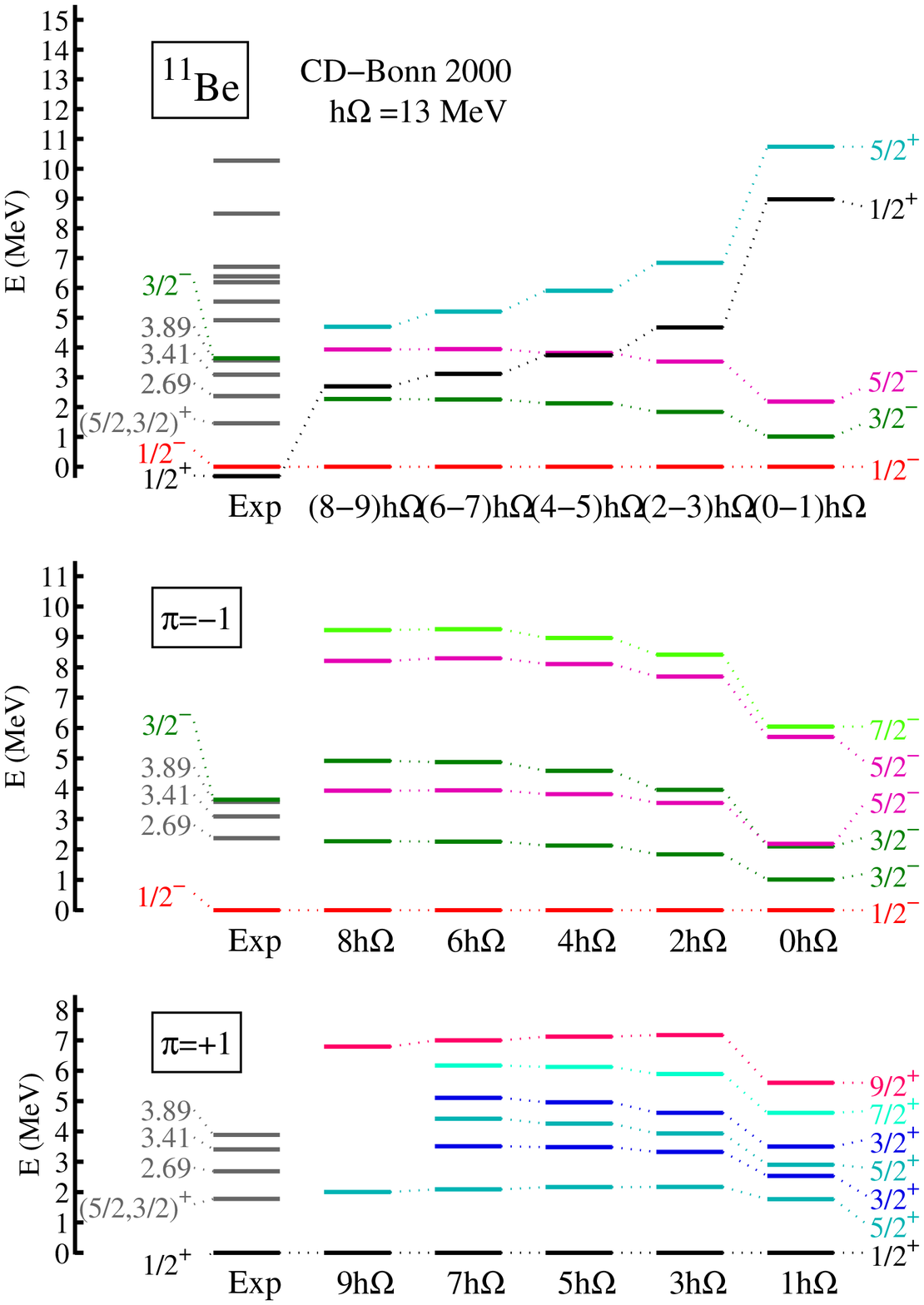}
  \caption{\co\ Excitation spectrum for \beel\ calculated using the
  \cdb\ interaction in 0\ho--9\ho\ model spaces with
  a fixed HO frequency of $\hbar\Omega = 13$~MeV. The experimental
  values are from Ref.~\cite{ajz90:506rev}. The two lower graphs show
  separately the negative- and positive-parity spectra, while the upper
  graph shows the combined spectrum with selected states.%
  \label{fig:11cdbspec}}
\end{figure}
\begin{figure}[hbtp]
  \includegraphics*[width=0.9\columnwidth]{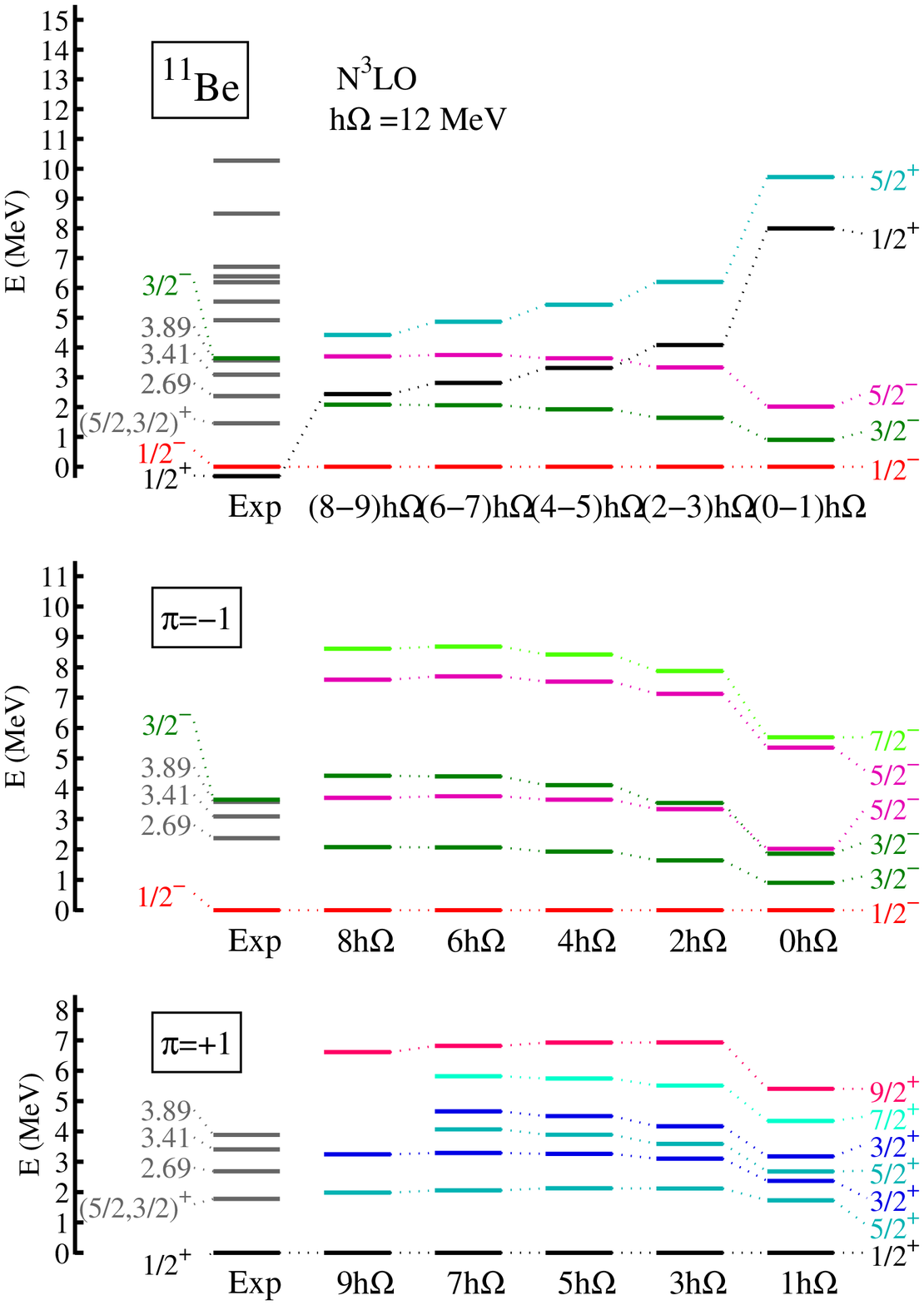}
  \caption{\co\ Excitation spectrum for \beel\ calculated using the
  \nlo\ interaction in 0\ho--9\ho\ model spaces with
  a fixed HO frequency of $\hbar\Omega = 12$~MeV. The experimental
  values are from Ref.~\cite{ajz90:506rev}. The two lower graphs show
  separately the negative- and positive-parity spectra, while the upper
  graph shows the combined spectrum with selected states.%
  \label{fig:11n3lospec}}
\end{figure}
\begin{figure}[hbtp]
  \includegraphics*[width=0.9\columnwidth]{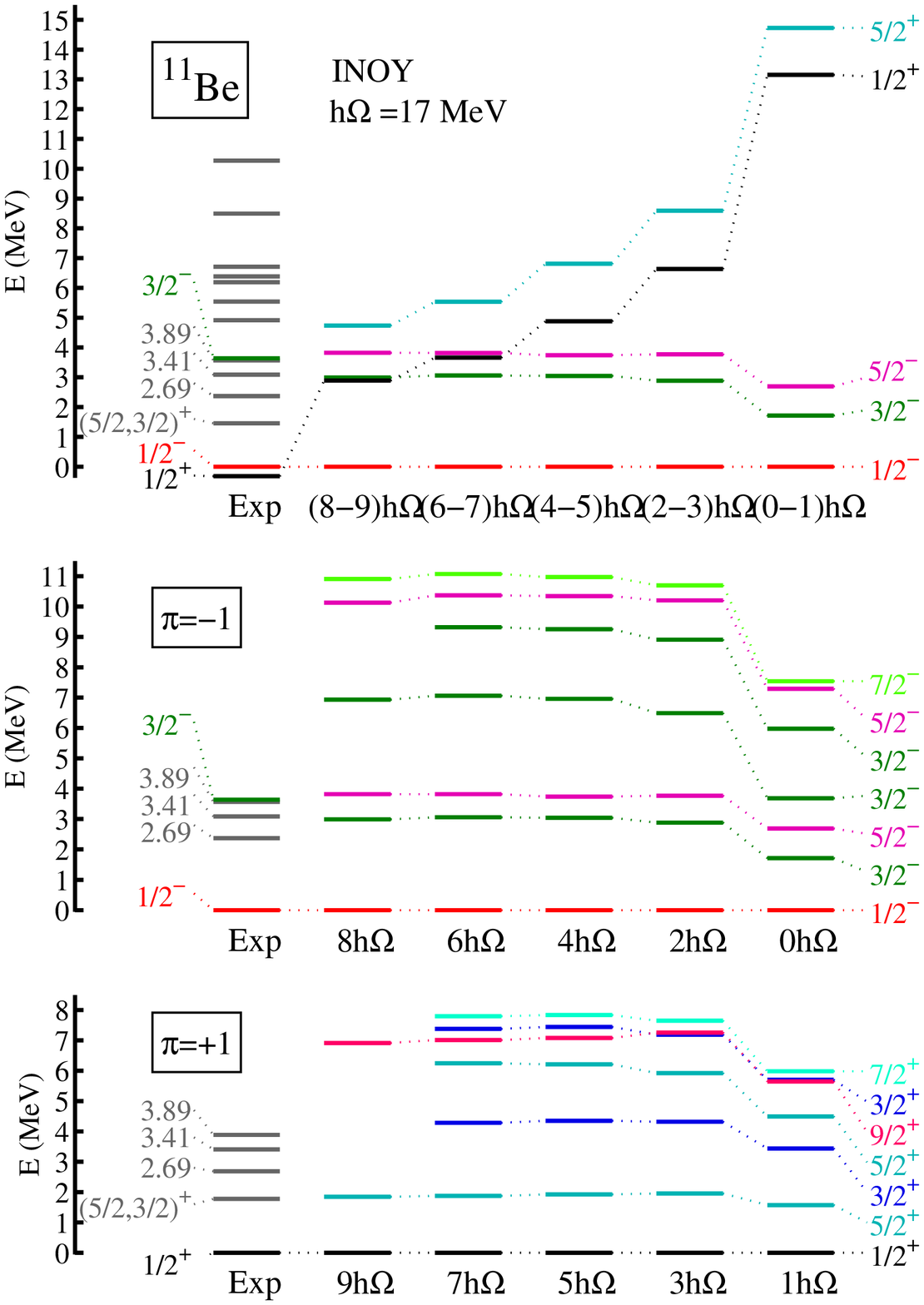}
  \caption{\co\ Excitation spectrum for \beel\ calculated using the
  \inoy\ interaction in 0\ho--9\ho\ model spaces with
  a fixed HO frequency of $\hbar\Omega = 17$~MeV. The experimental
  values are from Ref.~\cite{ajz90:506rev}. The two lower graphs show
  separately the negative- and positive-parity spectra, while the upper
  graph shows the combined spectrum with selected states.%
  \label{fig:11inoyspec}}
\end{figure}
\begin{table}[hbtp]
  \caption{Experimental and calculated energies (in [MeV]) of the lowest
    negative- and positive-parity states in \beel, as well as the
    magnetic moment (in [$\mu_N$]) of the ground state. Results for the
    \avp, \cdb, \nlo\, and \inoy\ \nn\ interactions are presented. These
    calculations were performed in the 8(9)$\ho$ model space for
    negative-(positive-)parity states, using the HO frequencies listed in
    Table~\ref{tab:optfreq}. $E_{x^-}$ denotes the excitation
    energy relative to the lowest negative-parity state. Experimental
    values are from~\cite{gei99:83,ajz90:506rev}. \vspace*{1ex}%
  \label{tab:11energies}}
  \begin{ruledtabular}
    \begin{tabular}{cccccc}
      \beel\  &        & \multicolumn{4}{c}{NCSM}                      \\   
              & Exp    & \inoy\    &  \cdb\    & \nlo\     & \avp\     \\
      \hline
      $E \jpt{1}{1}{-}{3}$
              & -65.16 & -62.40    & -56.95    & -56.57    & -55.52    \\ 
      $E\jpt{1}{1}{-}{3} - E_\mathrm{gs}$
              & 0.32   & -2.89     & -2.69     & -2.44     & -2.54     \\
      $E_{x^-} \jpt{1}{1}{-}{3}$
              & 0      & 0         & 0         & 0         & 0         \\
      $E_{x^-} \jpt{3}{1}{-}{3}$
              & ?\footnote[1]{There are large ambiguities in the
              experimental spin-parity assignments,
              cf. Table~\ref{tab:explevels}} 
                       & 2.99      & 2.27      & 2.08      & 2.04      \\
      $E_{x^-} \jpt{5}{1}{-}{3}$
	      & ?\footnotemark[1]
                       & 3.82      & 3.93      & 3.70      & 3.81      \\
      $E_{x^-} \jpt{3}{2}{-}{3}$ 
              & ?\footnotemark[1]
                       & 6.93      & 4.91      & 4.43      & 4.38      \\
      \hline
      $E_\mathrm{gs}\jpt{1}{1}{+}{3}$
              & -65.48 & -59.51    & -54.26    & -54.13    & -52.98    \\
      $E_x \jpt{1}{1}{+}{3}$
              & 0      & 0         & 0         & 0         & 0       \\
      $E_x \jpt{5}{1}{+}{3}$
              & ?\footnotemark[1]
                       & 1.85      & 2.01      & 1.98      & 2.03    \\
      \hline
      $\mu_\mathrm{gs}$
              & -1.6816(8)
                       & -1.47     & -1.55     & -1.58     & -1.58     \\
    \end{tabular}
  \end{ruledtabular}
\end{table}

The low-lying experimental spectrum is compared to our NCSM calculated
levels (obtained using the four different \nn\ interactions) in
Figs.~\ref{fig:11v8pspec}--\ref{fig:11inoyspec}
%
%
and in Table~\ref{tab:11energies}. In the two lower panels of these
figures we show, separately, the negative- and positive-parity spectra,
while the upper panel shows the combined spectrum with selected
states. Note that, those experimental levels for which there is an
uncertainty in the parity assignment, are included in all three
panels. There are clear signs of convergence with increasing model
space. However, as was also observed for \beni, the relative position of
the negative- and positive-parity spectra has clearly not converged, and
the latter is still moving down. The most dramatic drop is observed in
the \inoy\ spectrum, thus indicating the importance of a $3N$
force. With this particular interaction, the $1/2^+$ level actually ends
up below all, but one, of the negative-parity states in the (8--9)\ho\
calculation. We refer to Sec.~\ref{sec:pospar} for further discussions
on this topic.

We stress again that the relative level spacings, observed when plotting
negative- and positive-parity states separately, is remarkably
stable. Furthermore, the ordering of the first six(four) levels of
negative(positive) parity, is the same for all four potentials. This
calculated level ordering is summarized in
Table~\ref{tab:theorordering}.
\begin{table}[hbtp]
  \caption{NCSM observed ordering (from left to right) of \beel\
  negative- and positive-parity states (separately). Note that all four
  \nn\ interactions used in this study give the same ordering for the
  first six(four) negative-(positive-)parity states.
  %
  %
  \vspace*{1ex}%
  \label{tab:theorordering}}
  \begin{ruledtabular}
    \begin{tabular}{cccccc}
      \multicolumn{6}{c}{Negative parity}\\
      $\frac{1}{2}^-$ & $\frac{3}{2}^-$ & $\frac{5}{2}^-$ &
      $\frac{3}{2}^-$ & $\frac{3}{2}^-$ & $\frac{5}{2}^-$ \\[1ex]
    \hline
      \multicolumn{6}{c}{Positive parity}\\
      $\frac{1}{2}^+$ & $\frac{5}{2}^+$ & $\frac{3}{2}^+$ &
      $\frac{5}{2}^+$ & \multicolumn{2}{c}{} \\ 
   \end{tabular}
  \end{ruledtabular}
\end{table}
Our results can, therefore, provide input to help resolve the
uncertainties of the experimental spin-parity assignments
(cf.~Table~\ref{tab:explevels}). Note in particular that some
experiments suggest that there are three low-lying $3/2^-$ states. The
task to compute three levels with the same spin quickly becomes very
time consuming with increasing dimension, since it requires many Lanczos
iterations. Therefore, this third state was studied in two separate
runs, using only the \avp\ and \inoy\ potentials, and is included in
Figs.~\ref{fig:11v8pspec} and \ref{fig:11inoyspec} up to the 6\ho\ model
space. These calculations confirm the existence of three low-lying
$3/2^-$ levels, but they also stress the presence of a $5/2^-$ state
which is not completely consistent with
Refs.~\cite{liu90:42,aoi97:616,hir04:738}. However, we can not rule out
the possibility of a low-lying intruder 2\ho-dominated state, which
would avoid detection in our study. These states have a different
convergence pattern than 0\ho\ states and generally appear at too high
an excitation energy in the smaller model spaces, see
e.g. Ref.~\cite{cau01:64}.

In summary, our results suggest that there are two excited
positive-parity states below 4~MeV (rather than three as stated in
Ref.~\cite{ajz90:506rev}). The 1.78~MeV level should be a $5/2^+$ state,
while either the 3.41 or the 3.89~MeV level is a $3/2^+$. Our results do
not support the presence of a high-spin ($J \geq 7/2$) state, which one
can find in Ref.~\cite{ajz90:506rev}. We do observe three low-lying
$3/2^-$ states although they are accompanied by a $5/2^-$ state.

The strength of the electric dipole transition between the two bound
states in \beel\ is of fundamental importance. This is an observable
which has attracted much attention since it was first measured in
1971~\cite{han71:3}, and again in 1983~\cite{mil83:28}. The cited value
of 0.36 W.u. is still the strongest known transition between low-lying
states, and it has been attributed to the halo character of the
bound-state wave functions. Unfortunately, by working in a HO basis, we
suffer from an incorrect description of the long-range asymptotics, and
we would need an extremely large number of basis states in order to
reproduce the correct form. This shortcoming of the HO basis is
illustrated by the fact that we obtain a value for the E1 strength which
is 20 times too small (see Table~\ref{tab:radius}). When studying the
dependence of this value on the size of the model space, we observe an
almost linear increase, indicating that our result is far from
converged.  For the $\left\{(4-5)\:-\:(6-7)\:-\:(8-9)\right\}\ho$
sequence of model spaces, the \beel\ E1 strength, $B\left( \mathrm{E}1;
\frac{1}{2}_{_1}^- \to \frac{1}{2}_{_1}^+ \right)$, calculated with the
\avp\ interaction increases as:
$\left\{0.0054\:-\:0.0059\:-\:0.0065\right\}$~[$e^2$fm$^2$].  The
corresponding sequence of results for \beni\ is: $B \left( \mathrm{E}1;
\frac{1}{2}_{_1}^+ \to \frac{3}{2}_{_1}^- \right) =
\left\{0.029\:-\:0.031\:-\:0.033\right\}$~[$e^2$fm$^2$], which
demonstrates a similar increase. However, for this nucleus we note that,
in the largest model space, our calculated E1 strength is only off by a
factor of two compared to experiment. In addition, a consistent result
is found for the much weaker $\frac{5}{2}_{_1}^+ \to \frac{3}{2}_{_1}^-
$ E1 transition in \beni, where we also obtain a factor of two smaller
$B(\mathrm{E}1)$ than experiment. These results accentuates the
anomalous strength observed for \beel. A simple explanation for the
failure of HO calculations in the \beel\ case was given by Millener
\emph{et al}~\cite{mil83:28}. It was shown that there is a strong
cancellation in the calculated E1 transition amplitude due to the
insufficient description of the long-range asymptotics (see in
particular Tables IV and V in Ref.~\cite{mil83:28}). By simply replacing
their HO single-particle wave functions with solutions to the
Schr\"odinger equation with a Woods-Saxon potential, they found that the
magnitude of the neutron $1s_{1/2}0p_{1/2}$ single-particle matrix
element increased significantly so that the cancellation was removed.
Even though our multi-\ho\ calculations give a significant improvement
of the calculated E1 strengths as compared to their simple
(0--1)\ho\ model, the underlying problem is still present.

Another operator which is sensitive to the long-range behavior of the
wave function is the point-nucleon radius. However, even though no
operator renormalization has been applied, our results show a fair
stability with increasing model space, and they are in rather good
agreement with experimental findings for both \beni\ and \beel\ (see
Table~\ref{tab:radius}). It is probably safe to assume that the missing
part of the \beel\ matter radius originates mainly in an underestimation
of the point-neutron radius. One should also remember that the
experimental results for matter radii, in these light systems, are
highly model-dependent and are usually theoretically extracted from
measurements of the interaction cross section. In addition, we have also
calculated the radii of the first excited state. For both isotopes it is
found that the unnatural-parity state has a $~10$\% larger neutron
radius than the natural-parity one, probably due to a larger admixture
of $sd$-shell neutrons. Finally, the ground-state magnetic
moment of \beel\ has been measured~\cite{gei99:83} and we find a
reasonable agreement with our calculated value, see
Table~\ref{tab:11energies}.
\begin{table}[hbtp]
  \caption{Nuclear ground-state radii (in [fm]) and the E1 strengths (in
    [$e^2$fm$^2$]) for the strong ground-state transitions in \beni\ and
    \beel. The NCSM calculations were performed in the 8(9)$\ho$ model
    space for negative-(positive-)parity states using the \avp\
    interaction. The GFMC result for \beni, with the same
    interaction~\cite{pie02:66}, is shown for comparison. Experimental
    values are
    from~\cite{til04:745,ajz90:506rev,tan88:206,fri95:60}. \vspace*{1ex}%
  \label{tab:radius}}
  \begin{ruledtabular}
    \begin{tabular}{ccccccc}
              & \multicolumn{5}{c}{\beni\jpt{3}{1}{-}{1} } \\
              & & & & \multicolumn{2}{c}{$B ( \mathrm{E}1 )$} \\
              & $R_n$      & $R_p$        & $R_\mathrm{mat}$    & 
      $\frac{1}{2}_{_1}^+ \to \frac{3}{2}_{_1}^-$  &
      $\frac{5}{2}_{_1}^+ \to \frac{3}{2}_{_1}^-$ \\
      \hline							       
      Exp     &            & 2.39         & 2.45(1)\footnote{Interaction
        radius}                                                 
      & 0.061(25) & 0.0100(84) \\
      NCSM    & 2.40       & 2.27         & 2.34                
      & 0.033      & 0.0057 \\
      GFMC    & ---        & 2.41(1)      & ---                 & --- & --- \\
      \hline
      \hline
              & \multicolumn{5}{c}{\beel\jpt{1}{1}{+}{3} } \\
               & & & & \multicolumn{2}{c}{$B ( \mathrm{E}1 )$} \\
             & $R_n$      & $R_p$        & $R_\mathrm{mat}$   & 
      \multicolumn{2}{c}{$\frac{1}{2}_{_1}^- \to \frac{1}{2}_{_1}^+$} \\
      \hline							       
      Exp     &            &              & 2.86(4)            & 
      \multicolumn{2}{c}{0.116(12)} \\
      NCSM    & 2.66       & 2.30         & 2.54               & 
      \multicolumn{2}{c}{0.0065} \\
    \end{tabular}
  \end{ruledtabular}
\end{table}

\begin{figure}[hbtp]
  \includegraphics*[width=0.9\columnwidth]{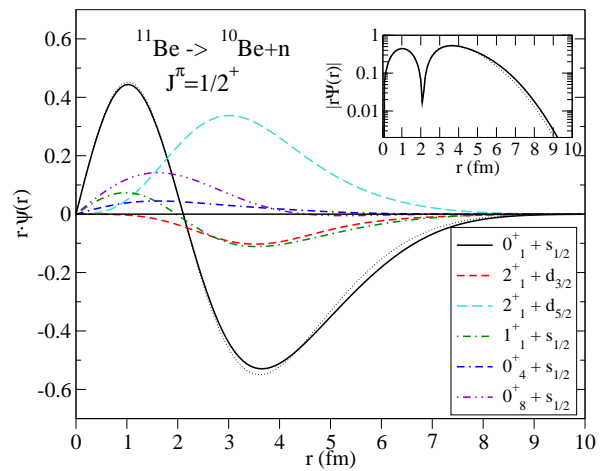}
  \caption{\co\ The largest radial overlap functions for the
    \beel\jpt{1}{1}{+}{3} state decomposed as $\nuc{10}{Be} + n$ in
    $jj$-coupling. The results presented here were obtained with the
    \cdb\ interaction ($\ho = 14$~MeV) with the \beel(\nuc{10}{Be}) wave
    function calculated in a $7(6)\hbar\Omega$ model space. The thin,
    dotted line shows the dominant overlap function calculated in a
    smaller $5(4)\hbar\Omega$ model space.
  \label{fig:11overlap}}
\end{figure}

The standard halo picture of the \beel\ ground state is a simple
two-body configuration consisting of an inert \nuc{10}{Be} core coupled
to an $s_{1/2}$ valence neutron. Theoretical estimates of the
spectroscopic factor for this component range from 0.55 to 0.92, see
e.g. Table~1 in Ref.~\cite{win01:683}. The experimental situation is
also unclear since the extracted results are generally
model-dependent. In the literature one can find values from 0.36 to 0.8,
see e.g. Fig.~8 in Ref.~\cite{pal03:68}. An important question is to
which extent the first-excited \nuc{10}{Be}$\left( 2_1^+ \right)$ state
contributes to the simple two-body configuration. The formalism for
investigating cluster structures of NCSM eigenstates was recently
developed in Ref.~\cite{nav04:70_2}. We have calculated the overlap of
the \beel\jpt{1}{1}{+}{3} state with different $\nuc{10}{Be} + n$
channels. To this aim, the \beel(\nuc{10}{Be}) wave functions were
calculated using the \cdb\ interaction in a 7(6)\ho\ model space. We
used a HO frequency of $\ho = 14$~MeV, which corresponds to the optimal
value for calculating binding energies in these two model spaces. The
largest overlap functions (in $jj$ coupling) are presented in
Fig.~\ref{fig:11overlap}, while the corresponding spectroscopic factors
(the overlap function squared and integrated over all $r$) are
summarized in Table~\ref{tab:11specfac}. Several additional channels,
such as the overlap with the second excited $2_2^+$ state in
\nuc{10}{Be}, were also computed but their spectroscopic factors were
found to be very small ($\lesssim 0.001$).
\begin{table}[hbtp]
  \caption{Spectroscopic factors for the \beel\ \jpt{1}{1}{+}{3} ground
    state decomposed as $\nuc{10}{Be} + n$ in $jj$
    coupling. The results presented here were obtained with the \cdb\
    interaction ($\ho = 14$~MeV) with the \beel(\nuc{10}{Be}) wave
    function calculated in a $7(6)\hbar\Omega$ model
    space. For comparison, we list spectroscopic factors extracted from
    three recent experiments utilizing different reactions.\vspace*{1ex}%
  \label{tab:11specfac}}
  \begin{ruledtabular}
    \begin{tabular}{cccccc}
      \multicolumn{2}{c}{$\nuc{10}{Be} \otimes n$} &
        & Transfer & Knockout &
        Breakup \\
	$J^\pi$ &  $(l,j)$ & NCSM
	& \cite{win01:683}\footnote{DWBA analysis of $\beel(p,d)$.}
        & \cite{aum00:84}\footnote{From $\beni \left(
        \beel,\nuc{10}{Be}+\gamma \right)$} 
	& \cite{pal03:68}\footnote{Spectroscopic factors extracted from
        \beel\ breakup on lead and carbon targets respectively.} \\
      \hline							
      $0_1^+$ & $\left( 0, \frac{1}{2} \right) $     & 0.818 
        & 0.67-0.80 & 0.78 & 0.61(5), 0.77(4) \\
      $2_1^+$ & $\left( 2, \frac{5}{2} \right) $     & 0.263 
        & 0.09-0.16 \\
      $2_1^+$ & $\left( 2, \frac{3}{2} \right) $     & 0.022 \\
      $1_1^+$ & $\left( 0, \frac{1}{2} \right) $     & 0.032 \\
      $0_4^+$ & $\left( 0, \frac{1}{2} \right) $     & 0.005 \\
      $0_8^+$ & $\left( 0, \frac{1}{2} \right) $     & 0.037 \\
    \end{tabular}
  \end{ruledtabular}
\end{table}
Several observations can be made when studying these results: (1) The
\beel\ ground state has a large overlap with $\left[ \nuc{10}{Be}(0_1^+)
\otimes n\left(s_{1/2}\right) \right]$ ($S= 0.82$), but also with the
core-excited $\left[ \nuc{10}{Be}(2_1^+) \otimes n\left(d_{5/2}\right)
\right]$ channel ($S=0.26$). These results are in good agreement with
the consensus of recent experimental studies, see
e.g. Refs.~\cite{aum00:84,win01:683}.  (2) The thin dotted line in
Fig.~\ref{fig:11overlap} shows the $\left[ \nuc{10}{Be}(0_1^+) \otimes
n\left(s_{1/2}\right) \right]$ overlap function calculated in a smaller
model space. From this comparison it is clear that the results are quite
stable with regards to a change in \nm. The interior part does hardly
change at all, while the tail is slowly extending towards larger
inter-cluster distances. This statement is true for all channels shown
in the figure except for those involving the two high-lying $0^+$ states
[see bullet (4) below]. (3) The inset shows the main component plotted
on a logarithmic scale. This graph clearly demonstrates the fact that
our HO basis is not large enough to reproduce the correct asymptotic
behavior. Even though the tail is extending further with increasing \nm,
it still does not reach the expected exponential decay. Instead it dies
of too fast. (4) Our calculated \nuc{10}{Be} $0_4^+$ and $0_8^+$ states
are found to be 2\ho\ dominated, and their binding energies have not
converged in the NCSM calculation. The cluster overlaps with these
states do not display the same stability as observed for the other
channels. Instead, there is a large dependence on \nm. A similar result
was found in Ref.~\cite{nav04:70_2} and it is just another manifestation
of the slower convergence of the 2\ho\ states in the NCSM.

Finally, we compare, in Tables~\ref{tab:11conf} and~\ref{tab:11spocc},
the resulting configurations and the occupancies of single-particles
states obtained with different interactions. Again, it is clear that the
\inoy\ eigenstates have a larger fraction of low-\ho\ excitations and
that this interaction results in a different single-particle spectrum
due, most likely, to a stronger spin-orbit interaction.
\begin{table}[hbtp]
  \caption{Calculated configurations of the first negative- and
    positive-parity states in \beel. Results obtained in our largest
    model spaces (8$\hbar\Omega$ and 9$\hbar\Omega$, respectively) are
    presented. The calculations were performed with the HO frequencies
    listed in Table~\ref{tab:optfreq}.\vspace*{1ex}%
  \label{tab:11conf}}
  \begin{ruledtabular}
    \begin{tabular}{lccccc}
      & \multicolumn{5}{c}{\beel\jpt{1}{1}{-}{3} (8$\hbar\Omega$ model space)}\\
      \nn\ interaction & 0$\hbar\Omega$ & 2$\hbar\Omega$ & 4$\hbar\Omega$
        & 6$\hbar\Omega$ & 8$\hbar\Omega$ \\
      \hline
      \inoy\ & 0.59 & 0.17 & 0.14 & 0.06 & 0.04 \\
      \cdb\  & 0.51 & 0.20 & 0.15 & 0.08 & 0.06 \\
      \nlo\  & 0.49 & 0.22 & 0.15 & 0.08 & 0.06 \\
      \avp\  & 0.48 & 0.21 & 0.16 & 0.08 & 0.07 \\[2ex]
      \hline
      & \multicolumn{5}{c}{\beel\jpt{1}{1}{+}{3} (9$\hbar\Omega$ model space)}\\
      \nn\ interaction & 1$\hbar\Omega$ & 3$\hbar\Omega$ & 5$\hbar\Omega$
        & 7$\hbar\Omega$ & 9$\hbar\Omega$ \\
      \hline
      \inoy\ & 0.56 & 0.20 & 0.14 & 0.06 & 0.04 \\
      \cdb\  & 0.50 & 0.21 & 0.15 & 0.08 & 0.06 \\
      \nlo\  & 0.49 & 0.22 & 0.16 & 0.08 & 0.06 \\
      \avp\  & 0.48 & 0.22 & 0.16 & 0.08 & 0.07 \\
    \end{tabular}
  \end{ruledtabular}
\end{table}
\begin{table}[hbtp]
  \caption{Calculated occupations of neutron single-particle levels for
    the first negative- and positive-parity states in \beel. Results
    obtained in our largest model spaces (8$\hbar\Omega$ and
    9$\hbar\Omega$, respectively) are presented. The calculations were
    performed with the HO frequencies listed in
    Table~\ref{tab:optfreq}.\vspace*{1ex}%
  \label{tab:11spocc}}
  \begin{ruledtabular}
    \begin{tabular}{lcccccc}
      & \multicolumn{5}{c}{\beel\jpt{1}{1}{-}{3} (8$\hbar\Omega$ model
      space)}\\
      \nn\ interaction & $0s_{1/2}$ & $0p_{1/2}$ & $0p_{3/2}$
        & $1s_{1/2}$ & $0d_{3/2}$ & $0d_{5/2}$ \\
      \hline
      \inoy\ & 1.862& 1.078& 3.643& 0.046& 0.065& 0.075\\
      \cdb\  & 1.835& 1.093& 3.597& 0.066& 0.062& 0.072\\
      \nlo\  & 1.832& 1.095& 3.586& 0.073& 0.062& 0.072\\
      \avp\  & 1.828& 1.094& 3.579& 0.073& 0.061& 0.072\\[2ex]
      \hline
      & \multicolumn{5}{c}{\beel\jpt{1}{1}{+}{3} (9$\hbar\Omega$ model
      space)}\\
      \nn\ interaction & $0s_{1/2}$ & $0p_{1/2}$ & $0p_{3/2}$
        & $1s_{1/2}$ & $0d_{3/2}$ & $0d_{5/2}$ \\
      \hline
      \inoy\ & 1.845 & 0.504& 3.300& 0.658& 0.086& 0.345\\
      \cdb\  & 1.824 & 0.600& 3.181& 0.742& 0.088& 0.285\\
      \nlo\  & 1.823 & 0.616& 3.153& 0.752& 0.091& 0.281 \\
      \avp\  & 1.820 & 0.630 & 3.135 & 0.768 & 0.088 & 0.265
      \\[2ex] 
    \end{tabular}
  \end{ruledtabular}
\end{table}
%
\subsection{\label{sec:pospar}Parity inversion}
\begin{figure}[hbtp]
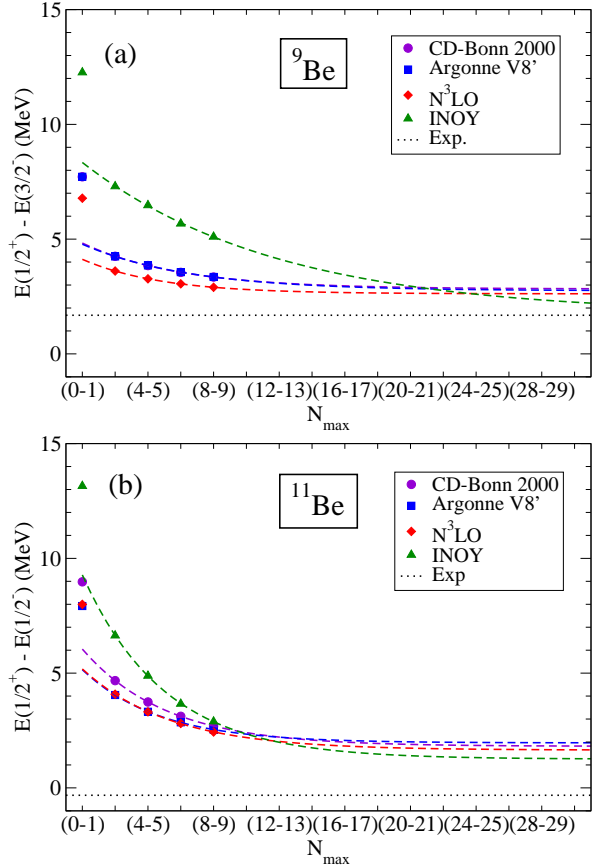

  \includegraphics*[width=0.9\columnwidth]{fig10a_a9energyminfixfreq_paper.eps}\\
  \includegraphics*[width=0.9\columnwidth]{fig10b_be11energyminfixfreq_paper.eps}
  \caption{\co\ Basis size dependence of the calculated $E_x\left(
    \frac{1}{2}_{_1}^{+} \right)$ excitation energy relative to the
    lowest negative-parity state in (a) \beni\ and (b) \beel. The results for
    four different \nn\ interactions are compared. For each potential, a
    single, fixed HO frequency was used (see
    Table~\ref{tab:optfreq}). The dashed lines correspond to exponential
    fits of the calculated data and, for illustration, these curves are
    extrapolated to larger model spaces.%
  \label{fig:beeminextrapol}}
\end{figure}

One of the main objectives of this study has been to investigate the
relative position of negative- and positive-parity states in the region
around \beel. As we have shown, none of our calculations reproduce the
parity inversion that is observed for this nucleus. However, considering
the slower convergence rate for 1\ho-dominated states in the NCSM, and
the large, but still finite, model spaces that we were able to use, our
results are actually very promising. In all nuclei, we found a fast drop
of the unnatural-parity states with respect to the natural ones. This
behavior has already been demonstrated in earlier NCSM studies, but the
drop that we observe in \beel\ is the most dramatic so far. Furthermore,
the results obtained with the \inoy\ interaction are clearly different
from the others, which indicates the significance that a realistic $3N$
force should have in a fundamental explanation of the parity
inversion. Note that \inoy\ is a two-body interaction, but that it
simulates the main effects of $3N$ forces by short-range, non-local
terms. Furthermore, the $^{3\!}P$ \nn\ interactions are slightly modified in
order to improve the description of $3N$ analyzing
powers. Fig.~\ref{fig:beeminextrapol} shows the calculated excitation
energy of the first positive-parity states in \beni\ and \beel\ as a
function of the basis size, \nm. For illustrative purposes we have
extrapolated our results to larger model spaces assuming an exponential
dependence on \nm, i.e.,~$E_x = E_{x,\,\infty} + a \exp\left( -b \nm
\right)$. Note that the (0--1)\ho\ points are excluded from the fits.
The extrapolated \inoy\ results end up below the other interactions; and
for \beni\ the curve is actually approaching the experimental
value. With all other interactions, the extrapolated excitation energy
is $\approx$1--2~MeV too high.

\begin{figure}[hbtp]
  \includegraphics*[width=0.9\columnwidth]{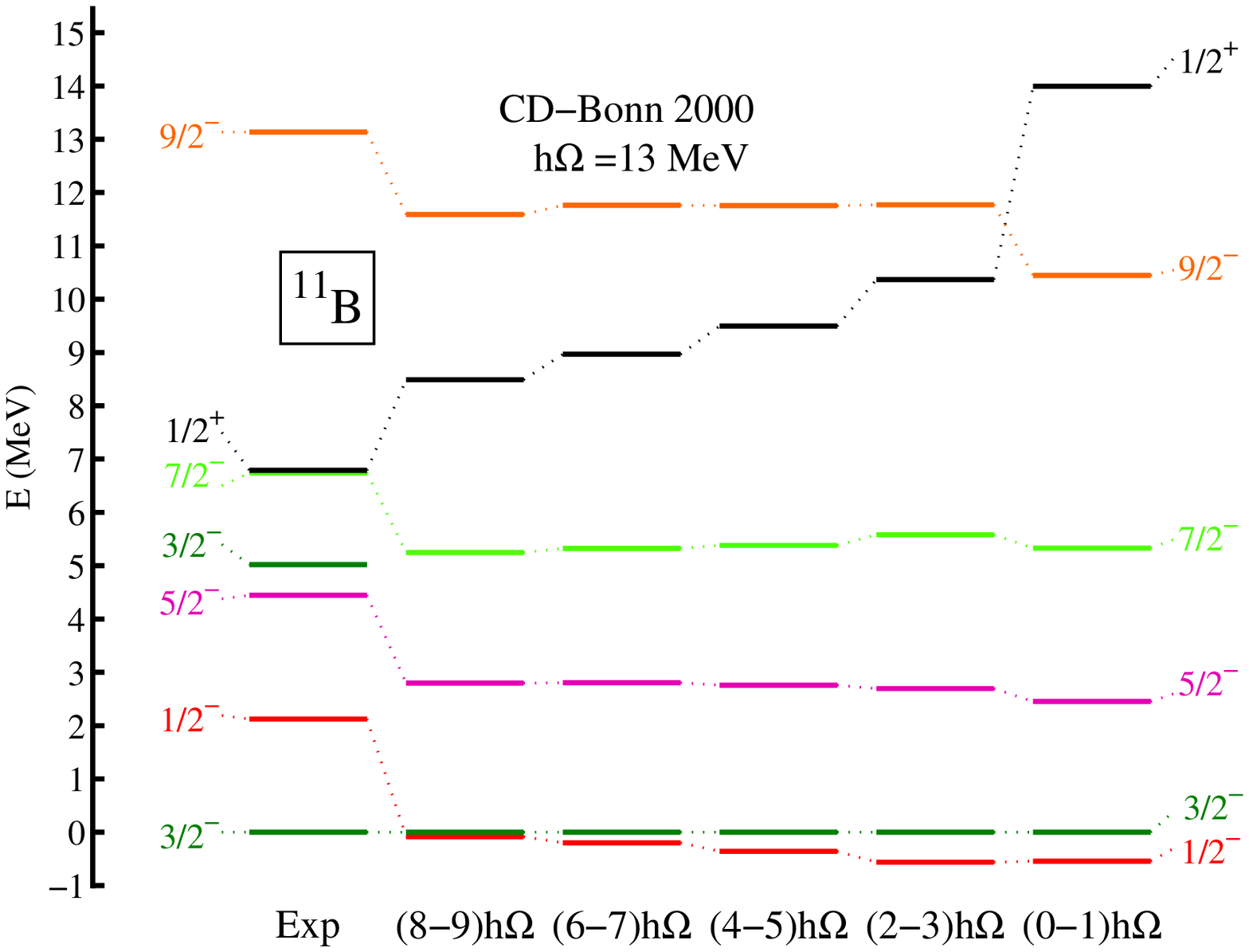}
  \caption{\co\ Excitation spectrum for \nuc{11}{B} calculated using the
  \cdb\ interaction in 0\ho--9\ho\ model spaces with a fixed HO
  frequency of $\hbar\Omega = 13$~MeV. The experimental levels are from
  Ref.~\cite{ajz90:506rev}. Note that there are many additional levels
  between the experimental $1/2^+$ and $9/2^-$ shown in the
  figure. However, we have only computed the first level for a given
  spin, and the $1/2^+$ was the only positive-parity state that was
  considered.%
  \label{fig:b11cdbspec}}
\end{figure}

When discussing the position of the first unnatural-parity state, it is
very interesting to study the systematics within the $A=11$ isobar and
the $N=7$ isotone. To this aim, we have performed large-basis
calculations for \nuc{11}{B} and \nuc{13}{C}. The diagonalization of the
\nuc{11}{B} Hamiltonian in the 9\ho\ space proved to be our largest
calculation so far. For \nuc{13}{C} we were only able to reach the 8\ho\
space. Both studies were performed using the \cdb\ interaction and an HO
frequency of $\ho = 13$~MeV. The ground-state binding energies (obtained
in the 8\ho\ space) are: $E \left( \nuc{11}{B}; \frac{3}{2}_{_1}^{-} \:
\frac{1}{2} \right) = 66.25$~MeV, and $E \left( \nuc{13}{C};
\frac{1}{2}_{_1}^{-} \: \frac{1}{2} \right) = 86.53$~MeV. Our calculated
\nuc{11}{B} spectrum, including the first negative-parity state for each
spin up to $J=9/2$, plus the lowest positive-parity state, is compared
to known experimental levels in Fig.~\ref{fig:b11cdbspec}. Note that we
obtain an incorrect $1/2^-$ ground-state spin in our largest model
space. However, the first $3/2^-$ and $1/2^-$ states are found to be
almost degenerate, and there is a trend indicating that the position of
these levels may eventually intersect as the basis size is increased. In
principle, a thorough frequency variation study should be performed in
order to clarify the fine details of the predictions. In any case, it is
clear that the level splitting is described incorrectly with this
interaction. Basically the same result was found in an earlier NCSM
study~\cite{nav03:68} using a three-body effective interaction derived
from \avp. In that paper, it was also shown that the correct level
ordering can be reproduced, and the splitting greatly improved, by
adding a realistic $3N$ force. For \nuc{13}{C} we have only computed the
lowest state for each parity. However, this nucleus has also been
studied previously using the NCSM. A spectrum obtained with the \cdb\
interaction was presented in Ref.~\cite{thi02:697}, while calculations
with a genuine $3N$ force were reported in Ref.~\cite{nav03:68}. In both
papers, the study was limited to negative-parity states and a smaller
model space (4\ho) was used.

\begin{figure}[hbtp]
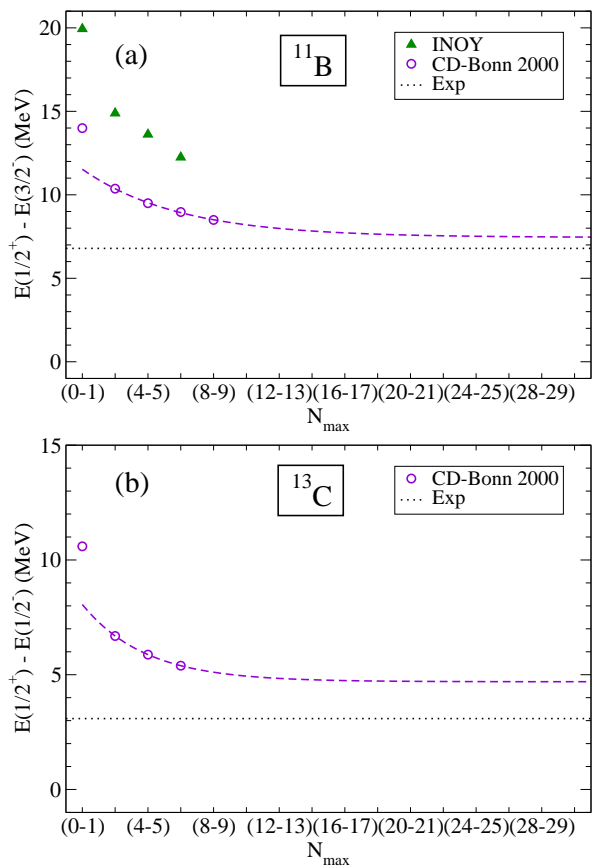

  \includegraphics*[width=0.9\columnwidth]
		   {fig12a_b11energyminfixfreq_paper.eps}\\
  \includegraphics*[width=0.9\columnwidth]
		   {fig12b_c13energyminfixfreq_paper.eps}
  \caption{\co\ Basis size dependence of the calculated $E_x\left(
    \frac{1}{2}_{_1}^{+} \right)$ excitation energy relative to the
    lowest negative-parity state in (a) \nuc{11}{B} and (b)
    \nuc{13}{C}. These results are obtained with the \cdb\ interaction
    and $\ho = 13$~MeV. The dashed lines correspond to exponential fits
    of the calculated data and, for illustration, these curves are
    extrapolated to larger model spaces.%
  \label{fig:bceminextrapol}}
\end{figure}

Let us now comment on our \nuc{11}{B} and \nuc{13}{C} results and return
to the important question of the position of the first positive-parity
state. The calculated $1/2^+$ excitation energy for these two nuclei, as
a function of \nm, is shown in Fig.~\ref{fig:bceminextrapol}. It is a
fascinating empirical fact that, by simply going from $Z=4 \to Z=6$, the
first $1/2^+$ state moves from being the ground state in \beel\ to
become an excited state at 3.1~MeV in \nuc{13}{C}. In the odd-$Z$
nucleus \nuc{11}{B}, the first positive-parity state is found quite high
in the excitation spectrum, namely at 6.8~MeV. It is a significant
success of the NCSM method, and of the \nn\ interactions being employed,
that these huge shifts are accurately reproduced in our
calculations. However, as can be seen from
Figs.~\ref{fig:beeminextrapol} and~\ref{fig:bceminextrapol}, the
calculated excitation energy always turns out to be too large. A
comparison of our extrapolated \cdb\ results shows that they exceed the
experimental values by $\approx$1--2~MeV for all four isotopes. As a
final remark, our \inoy\ results for \beni\ and \beel\ indicate that the
use of a realistic $3N$ force in a large basis space might correct this
discrepancy.
%
\section{\label{sec:conc}Conclusions}
We have performed large-basis \emph{ab initio} no-core shell model
calculations for \beni\ and \beel\ using four different realistic \nn\
interactions. One of these, the non-local \inoy\ interaction, has never
before been used in nuclear structure calculations. Although it is
formally a two-body potential, it reproduces not only \nn\ data (beware
of the fact that the $^{3\!}P$ interactions are slightly modified in the
IS-M version that we are using) but also the binding energies of
\nuc{3}{H} and \nuc{3}{He}. Therefore it has been of particular interest
for our current application, where we have striven to maximize the model
space by limiting ourselves to \nn\ interactions, but have still been
very much interested in the effects of three-body forces. We have
computed: binding energies, excited states of both parities,
electromagnetic moments and transition strengths, point-nucleon radii,
and also the core-plus-neutron cluster decomposition of the \beel\
ground state.

In summary, for the calculated spectra we found clear signs of
convergence, and a remarkable agreement between the predictions of
different \nn\ interactions. In particular, the relative level spacings
observed when plotting positive- and negative-parity states separately,
were found to be very stable and to agree well with experimental
spectra. This has allowed us to make some conclusions regarding the
largely unknown spin-parities of unbound, excited states in \beel. An
overall observation is that the \avp\ and \nlo\ potentials produce very
similar results, while \cdb\ gives slightly more binding. The \inoy\
interaction is clearly different; giving a much larger binding energy
and a stronger spin-orbit splitting. Both these effects would be
expected from a true $3N$ force, but are here achieved by the use of
short-range, non-local terms in the \nn\ interaction.

Furthermore, it was also clear from our study that our results for
observables connected to long-range operators, have not converged. These
calculations would clearly benefit from operator renormalization, in
order to correct for the limited model space being used. In particular,
the extremely strong E1 transition between the two bound states in
\beel, was underestimated by a factor of 20. We have discussed how this
illustrates the fact that the anomalous strength is due to the halo
character, and hence large overlap, of the initial and final state wave
functions; a property which is extremely hard to reproduce using a HO
basis. In the NCSM approach, there is no fitting to single-particle
properties, e.g., by the use of empirical interactions. Instead, the
effective interactions are derived from the underlying
inter-nucleon forces. Therefore, it is likely that a good
description of loosely bound, and unbound, single-particle states
might require a very large number of HO basis functions.

An important topic of this work has been the investigation of the parity
inversion found in \beel. We did not reproduce the anomalous $1/2^+$
ground state in our \emph{ab initio} approach, but did observe a
dramatic drop of the positive-parity excitation energies with increasing
model space. Furthermore, the behavior of our \inoy\ results suggests
that a realistic $3N$ force will have an important influence on the
parity inversion. However, in order to pursue this question further, an
improved computational capacity is needed. We have also performed
large-basis calculations for \nuc{11}{B} and \nuc{13}{C}. In this way,
we were able to put our \beel\ positive-parity results into a wider
context by studying the systematics within the $Z=4$ isotopes (\beni),
the $N=7$ isotone (\nuc{13}{C}), and the $A=11$ isobar
(\nuc{11}{B}). Although we found that the NCSM always overestimates the
excitation energy of the first unnatural-parity state, we did reproduce
the very large shifts observed for these different nuclei. This is an
important finding which leads us to the optimistic conclusion that the
parity-inversion problem should be possible to reproduce in the NCSM
starting from realistic inter-nucleon interactions.
%
\begin{acknowledgments}
This work was partly performed under the auspices of the
U. S. Department of Energy by the University of California, Lawrence
Livermore National Laboratory under contract No. W-7405-Eng-48. Support
from the LDRD contract No.~04--ERD--058, and from U.S.~Department of
Energy, Office of Science, (Work Proposal Number SCW0498) is
acknowledged.
\end{acknowledgments}
\bibliography{ncsm,f2b_refs,cf}
\end{document}